\documentclass[%
 reprint,
 amsmath,amssymb,
 aps,
floatfix,
]{revtex4-1}
\usepackage{natbib}
\usepackage{graphicx}
\usepackage{dcolumn}
\usepackage{bm}

\begin{document}


\title{Phase space analysis for a scalar-tensor model with kinetic and Gauss-Bonnet couplings}

\author{L. N. Granda}
 \email{luis.granda@correounivalle.edu.co}
\author{E. Loaiza}%
 \email{edwin.loaiza@univalle.edu.co}
  \altaffiliation[Also at ]{Universidad del Valle, Sede Buga.}
\affiliation{%
Departamento de Fisica, Universidad del Valle \\
A.A. 25360, Cali, Colombia
}%




\date{\today}

\begin{abstract}
We study the phase space for an scalar-tensor string inspired model of dark energy with non minimal kinetic and Gauss Bonnet couplings. The form of the scalar potential and of the coupling terms is of the exponential type, which give rise to appealing cosmological solutions. The critical points describe a variety of cosmological scenarios that go from matter or radiation dominated universe to dark energy dominated universe. There were found trajectories in the phase space departing from unstable or saddle fixed points and arriving to the stable scalar field dominated point corresponding to late-time accelerated expansion.

\begin{description}
\item[PACS numbers]
98.80.-k, 95.36.+x, 04.50.Kd
\end{description}
\end{abstract}

\pacs{Valid PACS appear here}
\maketitle


\section{\label{sec:Introduction}Introduction}
The explanation of late time acelerated expansion of the universe confirmed by different observational data \cite{Riess:1998cb}, \cite{Perlmutter:1997zf}, \cite{Kowalski:2008ez}, \cite{Hicken:2009dk}, \cite{Komatsu:2008hk}, \cite{Percival:2009xn}, represents one of the most important challenges of the modern cosmology. This phenomena is attributted to unknown kind of negative-pressure matter called dark energy (DE), which makes up about $70$\% of the total matter content of the universe. The current observational data are in agreement with the simplest posiblility of the cosmological constant being the source of DE, but there is no mechanism to explain its smallness (expressed in Planck units) in contradiction to the expected value as the vacuum energy in particle physics \cite{Peebles:2002gy}, \cite{Padmanabhan:2002ji}; and also observational data show a better fit for a redshift dependent equation of state. The modification of the energy-momentum tensor by introducing different kind of scalar field models, e.g., quintessence , tachyon, phantom, k-essence (for review see \cite{Copeland:2006wr, Sahni:2004ai, Padmanabhan:2002ji}), gives a dynamical interpretation to the DE, allowing the DE to pass through scalling behavior at early times, with ensuing transition to the late time accelerated phase and even to more exotic (quintom models) phantom phase, this last supported by the analysis of high redshift type Ia supernovae. From a pure geometrical point of view, the modified gravity theories, which are generalizations of the general relativity, have been widely considered to describe the early-time inflation and late-time acceleration without the introduction of any other dark component, and represent an important alternative to explain the dark energy (for review see \cite{Nojiri:2010wj}).
Despite the variety of DE models, it is however difficult to fulfil all observational requirements like the observed value of the equation of state parameter (EoS) of DE, $w\approx -1$ (or even the probably value $w<-1$), the current content of DE relative to that of dark matter (known as coincidence problem), and the estimated redshift transition between decelerated and accelerated phases, among others. To address some of these problems, several models include the coupling between DE and dark matter (DM), or introduce additional scalar fields with different attributes, etc. (see \cite{Copeland:2006wr} for a review).\\
Among the different alternatives to explain the dark energy, the scalar-tensor models which contain a direct coupling of the scalar field to the curvature, provide in principle a mechanism to evade the coincidence problem, allowing (in some cases) the crossing of the phantom barrier \cite{Perivolaropoulos:2005yv}, \cite{Fujii:2003pa}. The coupling of scalar field to curvature appears naturally after compatification of higher dimensional theories of gravity such as string theory, offering the possibility of connecting fundamental scalar fields with the nature of DE.\\
In the present work we consider a string and higher-dimensional gravity inspired scalar-tensor model, with non minimal kinetic and Gauss Bonnet (GB) couplings, to study late time cosmological dynamics. These terms are present in the next to leading $\alpha'$ corrections in the string effective action (where the coupling coefficients are functions of the scalar field) \cite{Metsaev:1987zx}, \cite{Meissner:1996sa} and have the notorious advantadge that lead to second order differential equations, preserving the theory ghost free. The GB term is topologically invariant in four dimensions, but nevertheless it affects the cosmological dynamics when it is coupled to a dynamically evolving scalar field through arbitrary function of the field. Some late time cosmological aspects of scalar field model with derivative couplings to curvature have been considered in \cite{Granda:2011eh}-\cite{Granda:2011zy}. On the other hand, the GB invariant coupled to scalar field has been extensively studied. In \cite{Nojiri:2005vv} the GB correction was proposed to study the dynamics of dark energy, where it was found that quintessence or phantom phase may occur in the late time universe. Different aspects of accelerating cosmologies with GB correction have been also discussed in \cite{Tsujikawa:2006ph}, \cite{Leith:2007bu}, \cite{Koivisto:2006xf}, \cite{Koivisto:2006ai}, \cite{Neupane:2006dp}. The modified GB theory applied to dark energy have been suggested in \cite{Nojiri:2005jg}, and further studies of the modified GB model applied to late time acceleration, have been considered among others, in  \cite{Neupane:2006dp}, \cite{Nojiri:2006je}, \cite{Cognola:2006eg}, \cite{Nojiri:2002hz}, \cite{Nojiri:2007te}. A model containing $\alpha'$ corrections to low-energy effective string action, has been considered in \cite{Cartier:2001is} to study the evolution of cosmological perturbations, where the final power spectra for scalar and tensor perturbations during inflation was obtained. In \cite{Granda:2011kx} solutions with Big Rip and Little Rip singularities have been considered, and in \cite{Granda:2011ja} the reconstruction of different cosmologial scenarios, including known phenomenologica models has been studied.\\
Scalar tensor theories are interesting alternatives to solve various issues of the DE problem, giving additionaly a relationship with the fundamental theories that could reveal itself in the current low-curvature universe (see \cite{Nojiri:2010wj} for review). In the present paper we combine the features of the GB coupling with those of the kinetic coupling to take the advantage of this string theory motivated model, in order to obtain viable late time cosmological dynamics. We consider the autonomus system for this model and study the cosmological implications coming out from the different critical points. The results of the model \cite{Koivisto:2006xf} with GB coupling have been reproduced at the limit when the kinetic coupling is negelcted; then the limit when the GB coupling is neglected (which corresponds to the model \cite{Granda:2009fh}, \cite{Granda:2010hb}) is considered, and finally the general case which combines the effects of both couplings is studied.
In section II we introduce the model and give the general equations, which are then expanded on the FRW metric. In section III we introduce the dynamical variables, solve the equations for the critical points and give an analysis of the different critical points. Some remarks are given in section IV.

\section{\label{sec:Dynamical Equations}Dynamical Equations }
We consider the following action which adds to the Gauss Bonnet coupling a kinetic coupling to curvature, which are present in the leading $\alpha'$ correction to the string effective action \cite{Meissner:1996sa}. 
\begin{equation}\label{eq1}
\begin{aligned}
S=&\int d^{4}x\sqrt{-g}\Big[\frac{1}{16\pi G} R +F_1(\phi)G_{\mu\nu}\partial^{\mu}\phi\partial^{\nu}\phi\\
&-\frac{1}{2}\partial_{\mu}\phi\partial^{\mu}\phi- V(\phi)-F_2(\phi){\cal G}\Big]+S_m
\end{aligned} 
\end{equation}
where $G_{\mu\nu}=R_{\mu\nu}-\frac{1}{2}g_{\mu\nu}R$, ${\cal G}$ is the 4-dimensional GB invariant ${\cal G}=R^2-4R_{\mu\nu}R^{\mu\nu}+R_{\mu\nu\rho\sigma}R^{\mu\nu\rho\sigma}$ and $S_m$ is the action for the matter, which includes the usual barionic matter and dark matter. The coupling $F_1(\phi)$ has dimension of $(length)^2$, and the coupling $F_2(\phi)$ is dimensionless. 
Nothe that compared with the more general action that leads to the second-order equations of motion (in metric and scalar field) \cite{Cartier:2001is}, we are neglecting derivative terms that are not directly coupled to curvature, of the form $\Box\phi\partial_{\mu}\phi\partial^{\mu}\phi$ and $(\partial_{\mu}\phi\partial^{\mu}\phi)^2$, which is is acceptable in a cosmological scenario with accelerated expansion. The properties of the GB invariant guarantee the absence of ghost terms in the theory. Hence, the equations derived from this action contain only second derivatives of the metric and the scalar field.\\
By varying Eq. (\ref{eq1}) with respect to metric we derive the gravitational field equations given by the expressions
\begin{equation}\label{eq2}
R_{\mu\nu}-\frac{1}{2}g_{\mu\nu}R=\kappa^2 \left(T_{\mu\nu}+T^{(m)}_{\mu\nu}\right)
\end{equation}
where $\kappa^2=8\pi G$, $T_{\mu\nu}^m$ is the usual energy-momentum tensor for the matter component, the tensor $T_{\mu\nu}$ represents the variation of the terms which depend on the scalar field $\phi$ and can be written as
\begin{equation}\label{eq3}
T_{\mu\nu}=T_{\mu\nu}^{\phi}+T_{\mu\nu}^{K}+T_{\mu\nu}^{GB}   
\end{equation}

where $T_{\mu\nu}^{\phi}$, correspond to the variations of the standard minimally coupled terms, $T_{\mu\nu}^{K}$ comes from the kinetic coupling, and $T_{\mu\nu}^{GB}$ comes from the variation of the coupling with GB. Due to the kinetic coupling with curvature and the GB coupling, the quantities derived from this energy-momentum tensors will be considered as effective ones. The respective components of the energy-momentum tensor (\ref{eq3}) are given by 
\begin{equation}\label{eq4}
T_{\mu\nu}^{\phi}=\nabla_{\mu}\phi\nabla_{\nu}\phi-\frac{1}{2}g_{\mu\nu}\nabla_{\lambda}\phi\nabla^{\lambda}\phi
-g_{\mu\nu}V(\phi)
\end{equation}
\begin{equation}\label{eq5}
\begin{aligned}
T_{\mu\nu}^{K}&=\left(R_{\mu\nu}-\frac{1}{2}g_{\mu\nu}R\right)F_1(\phi)\nabla_{\lambda}\phi\nabla^{\lambda}\phi\\
&+g_{\mu\nu}\nabla_{\lambda}\nabla^{\lambda}\left(F_1(\phi)\nabla_{\gamma}\phi\nabla^{\gamma}\phi\right)+R F_1(\phi)\nabla_{\mu}\phi\nabla_{\nu}\phi\\&-\frac{1}{2}(\nabla_{\mu}\nabla_{\nu}+\nabla_{\nu}\nabla_{\mu})\left(F_1(\phi)\nabla_{\lambda}\phi\nabla^{\lambda}\phi\right)\\
& -2F_1(\phi)\left(R_{\mu\lambda}\nabla^{\lambda}\phi\nabla_{\nu}\phi+R_{\nu\lambda}\nabla^{\lambda}\phi\nabla_{\mu}\phi\right)\\
&+g_{\mu\nu}R_{\lambda\gamma}F_1(\phi)\nabla^{\lambda}\phi\nabla^{\gamma}\phi+\nabla_{\lambda}\nabla_{\mu}\left(F_1(\phi)\nabla^{\lambda}\phi\nabla_{\nu}\phi\right)\\
&+\nabla_{\lambda}\nabla_{\nu}\left(F_1(\phi)\nabla^{\lambda}\phi\nabla_{\mu}\phi\right)-\nabla_{\lambda}\nabla^{\lambda}\left(F_1(\phi)\nabla_{\mu}\phi\nabla_{\nu}\phi\right)
\\
&-g_{\mu\nu}\nabla_{\lambda}\nabla_{\gamma}\left(F_1(\phi)\nabla^{\lambda}\phi\nabla^{\gamma}\phi\right)
\end{aligned}
\end{equation}
and 
\begin{equation}\label{eq6}
\begin{aligned}
T_{\mu\nu}^{GB}=&4\Big([\nabla_{\mu}\nabla_{\nu}F_2(\phi)]R
-g_{\mu\nu}[\nabla_{\rho}\nabla^{\rho}F_2(\phi)]R\\
&-2[\nabla^{\rho}\nabla_{\mu}F_2(\phi)]R_{\nu\rho}-2[\nabla^{\rho}\nabla_{\nu}F_2(\phi)]R_{\nu\rho}\\
&+2[\nabla_{\rho}\nabla^{\rho}F_2(\phi)]R_{\mu\nu}+2g_{\mu\nu}[\nabla^{\rho}\nabla^{\sigma}F_2(\phi)]R_{\rho\sigma}\\
&-2[\nabla^{\rho}\nabla^{\sigma}F_2(\phi)]R_{\mu\rho\nu\sigma}\Big)
\end{aligned}
\end{equation}
In this last expression the properties of the 4-dimensional GB invariant have been used (see \cite{Nojiri:2005vv}, \cite{Farhoudi:1995rc}).
By varying with respect to the scalar field gives the equation of motion
\begin{equation}
\label{eq7}
\begin{aligned}
&-\frac{1}{\sqrt{-g}}\partial_{\mu}\left[\sqrt{-g}\left(R F_1(\phi)\partial^{\mu}\phi-2R^{\mu\nu}F_1(\phi)\partial_{\nu}\phi+\partial^{\mu}\phi\right)\right]\\
&+\frac{dV}{d\phi}+\frac{dF_1}{d\phi}\left(R\partial_{\mu}\phi\partial^{\mu}\phi-2 R_{\mu\nu}\partial^{\mu}\phi\partial^{\nu}\phi\right)-\frac{dF_2}{d\phi}{\cal G}=0
\end{aligned}
\end{equation}
Let us consider the spatially-flat Friedmann-Robertson-Walker (FRW) metric,
\begin{equation}\label{eq8}
ds^2=-dt^2+a(t)^2\left(dr^2+r^2d\Omega^2\right)
\end{equation}
where $a(t)$ is the scale factor. Replacing this metric in Eqs. (\ref{eq2})-(\ref{eq8}) we obtain the set of equations describing the dynamical evolution of the FRW background and the scalar field in the present model:
\begin{equation}\label{eq9}
H^2=\frac{\kappa^2}{3}\left(\rho_{DE}+\rho_m\right)
\end{equation}
\begin{equation}\label{eq10}
-2\dot{H}-3H^2=\kappa^2\left(p_{DE}+p_m\right)
\end{equation}
where
\begin{equation}\label{eq11}
\rho_{DE}=\frac{1}{2}\dot{\phi}^2+V(\phi)+9 H^2F_1(\phi)\dot{\phi}^2+24H^3\frac{dF_2}{d\phi}\dot{\phi}
\end{equation}
and
\begin{equation}\label{eq12}
\begin{aligned}
&p_{DE}=\kappa^2\Big[\frac{1}{2}\dot{\phi}^2-V(\phi)-\left(3H^2+2\dot{H}\right)F_1(\phi)\dot{\phi}^2\\&-2 H\left(2F_1(\phi)\dot{\phi}\ddot{\phi}+\frac{dF_1}{d\phi}\dot{\phi}^3\right)
-8H^2\frac{dF_2}{d\phi}\ddot{\phi}\\&
-8H^2\frac{d^2F_2}{d\phi^2}\dot{\phi}^2
-16H\dot{H}\frac{dF_2}{d\phi}\dot{\phi}-16H^3\frac{dF_2}{d\phi}\dot{\phi}\Big]
\end{aligned}
\end{equation}

\begin{equation}\label{eq13}
\begin{aligned}
&0=\ddot{\phi}+18 H^3F_1(\phi)\dot{\phi}+12 H\dot{H}F_1(\phi)\dot{\phi}+3H\dot{\phi}+\frac{dV}{d\phi}\\
&+3 H^2\left(2F_1(\phi)\ddot{\phi}+\frac{dF_1}{d\phi}\dot{\phi}^2\right)+24\left(\dot{H}H^2+H^4\right)\frac{dF_2}{d\phi}
\end{aligned}
\end{equation}
In the present study we assume that the matter sector is modeled by an ideal fluid obeying the equation of state (EoS) $p_m=w_m\rho_m$ with constant equation of state parameter $w_m$(i.e. mostly non relativistic matter with $p_m=0$), whose energy density satisfies the usual continuity equation $\dot{\rho_m}+3H(\rho_m+p_m)=0$. 
Based on effective limits of fundamental theories like supergravity or string theory, the kinetic and GB couplings become exponentials of the scalar field (in leading $\alpha'$ correction in the case of string theory), but if take into consideration higher order corrections in $\alpha'$ expansion in the effective string theory, the couplings should change. In general the couplings $F_1(\phi)$ and $F_2(\phi)$ could be arbitrary functions of the scalar field, which gives more general character to the model (\ref{eq1}) where the couplings should be constrained by known observational limits, but allow to increase the number of phenomenologically viable solutions to the DE problem in comparison to the simple exponentials. In the present work we will adopt the following exponential form for the couplings and the potential
\begin{equation}\label{eq14}
\begin{aligned}
F_1(\phi)&=F_{10} e^{\alpha\kappa\phi/\sqrt{2}}\\ F_2(\phi)&=F_{20} e^{\beta\kappa\phi/\sqrt{2}}\\
V(\phi)&=V_0 e^{-\lambda\kappa\phi/\sqrt{2}}
\end{aligned}
\end{equation}
In the case of the model (\ref{eq1}) without potential, the exponential couplings lead to the important power-law evolution as shown in \cite{Granda:2011ja}, but in that case the exponen\-tial behavior appeared from reasonable restrictions on the relative densities corresponding to kinetic and GB couplings. \\
In order to perform the dynamical system analysis, let's introduce the following dimensionless variables:
\begin{equation}\label{eq15}
\begin{aligned}
&x=\frac{\kappa\dot{\phi}}{\sqrt{2}H} 
&y=\frac{\kappa^2 V}{H^2}\,\,\,\,\,
&k=3\kappa^2F_1\dot{\phi}^2 \\
&g=8\kappa^2 H\dot{\phi}\frac{dF_2}{d\phi} &\Omega_m=\frac{\kappa^2\rho_m}{3H^2}\,\, &\epsilon=\frac{\dot{H}}{H^2}
\end{aligned}
\end{equation}
In fact, these variables are related with the relative density parameters defined for the different sectors of the model
\begin{equation}\label{eq16}
\begin{aligned}
\Omega_{\phi}&=\frac{\kappa^2\rho_{\phi}}{3H^2}=\frac{1}{3}(x^2+y),\,\,\,\, \Omega_k=\frac{\kappa^2\rho_k}{3H^2}=k,\\ \Omega_{GB}&=\frac{\kappa^2\rho_{GB}}{3H^2}=g
\end{aligned}
\end{equation}
where
\begin{equation}\label{eq17}
\begin{aligned}
\rho_{\phi}&=\frac{1}{2}(\dot{\phi}^2+V),\,\,\,\, \rho_k=9\kappa^2H^2F_1\dot{\phi}^2,\\ \rho_{GB}&=24\kappa^2H^3\frac{dF_2}{dt}
\end{aligned}
\end{equation}
The Eq.(\ref{eq9}) imply the following restriction on the density parameters
\begin{equation}\label{eq18}
\Omega_{\phi}+\Omega_k+\Omega_{GB}+\Omega_m=1
\end{equation}
according to the expression for $\rho_{\phi}$ we can define the equation of state for the uncoupled terms of the scalar field (which satisfiy the continuity equation in the limit $F_1, F_2\rightarrow 0$) as
\begin{equation}\label{eq18a}
w_{\phi}=\frac{x^2-y}{x^2+y}
\end{equation}
and in general, the effective equation of state (EoS) can be written as
\begin{equation}\label{eq19}
\begin{aligned}
w_{eff}&=w_{\phi}\Omega_{\phi}+w_k\Omega_k+w_{GB}\Omega_{GB}+w_m\Omega_m\\
&=-1-\frac{2}{3}\epsilon
\end{aligned}
\end{equation}

Introducing the e-folding variable $N=\log a$, in terms of the variables (\ref{eq15}), the evolution equations (\ref{eq9})-(\ref{eq13}) can be transformed into the following first-order autonomus system
\begin{equation}\label{eq20}
x^2+y+3k+3g+3\Omega_m-3=0
\end{equation}
\begin{equation}\label{eq21}
2x x'+2(3+\epsilon)x^2+y'+2\epsilon y+k'+2(3+2\epsilon)k+3(1+\epsilon)g=0
\end{equation}
\begin{equation}\label{eq22}
\begin{aligned}
&2\epsilon +3+x^2-y-\frac{2}{3}k'-\frac{1}{3}(3+2\epsilon)k\\
&-g'-(2+\epsilon)g+w_m(3-x^2-y-3k-3g)=0
\end{aligned}
\end{equation}

\begin{equation}\label{eq24}
y'=-(\lambda x+2\epsilon)y
\end{equation}
\begin{equation}\label{eq25}
k'=(\alpha x+2\epsilon)k+2\frac{x'}{x}k
\end{equation}
\begin{equation}\label{eq26}
g'=(\beta x+2\epsilon)g+\frac{x'}{x}g
\end{equation}
where $'$ denotes derivative with respect to $N$. Note that the last three Eqs. come from the explicit form of the potential and the couplings given in (\ref{eq14}), and from Eq. (\ref{eq22}), which comes from Eqs. (\ref{eq10}) and (\ref{eq12}), follows the expression for the slow-roll parameter $\epsilon$
\begin{equation}\label{eq27}
\epsilon=\frac{9+3(x^2-y-k-g')-2k'-6g+9w_m\Omega_m}{2k+3g-6}
\end{equation}
See appendix A for details on $\epsilon$ in terms of the dynami\-cal variables and equations for the critical points.
There are three interesting cases of this dynamical system that can be considered separately.
\section{\label{sec:The critical point} The critical points}

The critical or fixed points of the autonomous system (\ref{eq20})-(\ref{eq26}) are the solutions to the equations $x'=0$, $y'=0$, $k'=0$ and $g'=0$. These critical points represent interesting cosmological solutions since they lead to scaling behavior of the dark energy component, in which the scalar field mimics the background fluid energy density that describes radiation or matter. There are also scalar field dominated fixed points that correspond to power-law accelerated expansion, which is relevant for the late time universe. 
The coordinates of the fixed points may be used to analyze the behavior of the model at the fixed points themselves. Thus, the fixed points $(x_c,y_c,k_c,g_c)$ of the system (\ref{eq20})-(\ref{eq26}) lead to the following solutions for the scale factor and the scalar field: from the first and last equations in (\ref{eq15}) one finds (ignoring the integration constants)
\begin{equation}\label{eq27a}
\phi=\frac{\sqrt{6}}{\kappa}x_c\ln a
\end{equation}
and 
\begin{equation}\label{eq27b}
H=-\frac{1}{\epsilon_c t}
\end{equation}
where $\epsilon_c$ is given by Eq. (\ref{eq27}) evaluated at the fixed point, and after using the constraint (\ref{eq20}) takes the form
\begin{equation}\label{eq27c}
\begin{aligned}
&\epsilon_c=\frac{1}{2k_c+3g_c-6}\Big(9+3(x_c^2-y_c-k_c-2g_c)\\&+3w_m(3-x_c^2-y_c-3k_c-3g_c)\Big)
\end{aligned}
\end{equation}
Integrating the equation (\ref{eq27b}) gives the known power-law behavior for the scale factor
\begin{equation}\label{eq27d}
a(t)\propto t^p
\end{equation}
with $p$ given by $p=-1/\epsilon_c$, and
\begin{equation}\label{eq27e}
\phi(t)=\frac{\sqrt{6}}{\kappa}x_c p\ln t
\end{equation}
Thus, the fixed points of the system  (\ref{eq20})-(\ref{eq26}) lead to power-law solutions for the scale factor and  logarithmic dependence of the scalar field with respect to the cosmic time. In order to analyze the stability of the solutions and find whether the system approaches one of the fixed points or not, we consider small perturbations $(\delta x,\delta y,...)$ around the critical points. This leads to the first order differential equations
\begin{equation}\label{eq27f}
\frac{d}{dN}\begin{pmatrix} \delta x\\ \delta y\\.\\. \end{pmatrix}={\cal A}\begin{pmatrix} \delta x\\ \delta y\\.\\. \end{pmatrix}
\end{equation}
where the matrix ${\cal A}$ is evaluated at the critical point and is given by
\begin{equation}\label{eq27g}
{\cal A}=\begin{pmatrix} \frac{\partial x'}{\partial x} & \frac{\partial x'}{\partial y} & . & .\\ \frac{\partial y'}{\partial x}& \frac{\partial y'}{\partial y} & . &.\\. &. &. &.\\ . &. &. &. \end{pmatrix}_{(x_c,y_c,.,.)}
\end{equation}
using the eigenvalues of the matrix ${\cal A}$ we can write the solutions for the linear perturbations as 
\begin{equation}\label{eq27h}
\begin{aligned}
&\delta x=c_1 e^{\lambda_1 N}+c_2 e^{\lambda_2 N}+...\\
&\delta y=d_1 e^{\lambda_1 N}+c_2 e^{\lambda_2 N}+...\\
&. \hspace{0.4cm} . \hspace{0.4cm} . \hspace{0.4cm} . \hspace{0.4cm} .  \hspace{0.4cm}.  \hspace{0.4cm}.\\
\end{aligned}
\end{equation}
where $\lambda_1, \lambda_2, ... $ are the eigenvalues corresponding to the fixed point $(x_c, y_c, ...)$ and $c_1, c_2, ...$ are the integration constants. As can be inferred from the behavior of the exponentials in the solution (\ref{eq27h}), one can classify the fixed points according to their stability properties as follows:\\
{\bf Stable}: if all the eigenvalues $\lambda_1, \lambda_2,....$ are real negative.\\
{\bf Unstable}: if all the eigenvalues $\lambda_1, \lambda_2, ...$ are real positive.\\
{\bf Saddle}: if there is at least one eigenvalue with opposite sign with respect to the others.\\
{\bf Stable spiral}: if the real parts of the eigenvalues are negative.\\
The fixed points classified as stable or stable spirals are also called attractors. Note that the linear perturbations around the attractor are exponentially damped. As long as the solution is an attractor, then independently of the initial conditions (or for a wide range of initial conditions) the scalar field finally enters in the scaling regime, contributing to the solution of the fine-tuning problem of the dark energy.

We will consider the model with each of the couplings separately, and then the model with all the couplings, in order to understand and compare the role of the couplings, specially in late time cosmology.

\subsection{\label{sec:GBL}The Gauss-Bonnet coupling (limit $k=0$)}
This limit takes place when the kinetic coupling is absent, which has been already considered in \cite{Koivisto:2006ai, Koivisto:2006xf}. This case is obtained from the system (\ref{eq20})-(\ref{eq26}) by making $k=0$ and $\alpha=0$. There are two fixed points co\-rres\-pon\-ding to scaling solutions, one of them co\-rres\-pon\-ding to stable spiral. The critical points are:\\

\noindent {\bf GB1}: $(x_c,y_c,g_c)=(0,0,1)$,\hspace{1cm} with eigenvalues $[2,2,-(1+3\,w_m)]$. The density parameters take the values: $\Omega_m=0$,  $\Omega_{GB}=1$, in this Gauss-Bonnet dominated critical point  $w_{eff}=-1/3$ which co\-rres\-ponds to the divide between decelerated and accelerated ex\-pan\-sion. This is a saddle fixed point for $w_m>-1/3$.\\

\noindent {\bf GB$2_{\pm}$}: $(x_c,y_c,g_c)=(\pm\sqrt{3},0,0)$, with eigenvalues $[3(1-w_m),-6\pm\sqrt{3}\,\beta,6\mp \sqrt{3}\,\lambda]$. The main den\-si\-ties and EoS are $\Omega_m=0$, $\Omega_{\phi}=0$, $w_{eff}=1$. This critical point is dominated by the kinetic term and leads to the equation of state characteristic of ``stiff matter''. The solution with $\dot{\phi}>0$ is a saddle point whenever $w_m<1$, $\beta<2\,\sqrt{3}$, $\lambda>2\,\sqrt{3}$ or for $\lambda,\beta<2\,\sqrt{3}$ or $\lambda,\beta> 2\,\sqrt{3}$.\\

\begin{widetext}

\begin{figure}[h]
    \centering
    \includegraphics [scale=0.5]{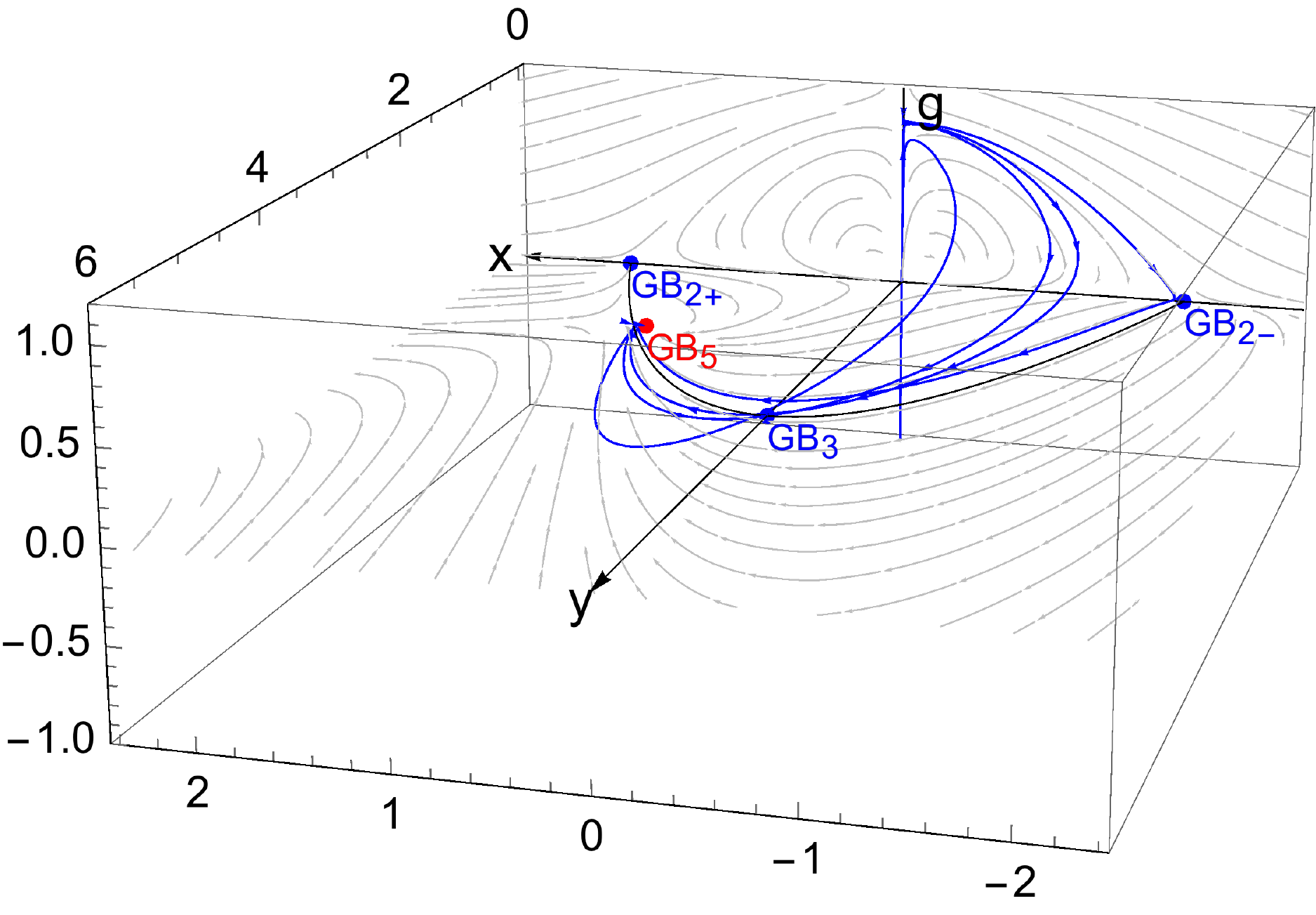}
    \caption{The phase portrait of the model for the values of the parameters ($\beta=1$, $\lambda=\frac{81}{32}$, $w_m=0$) showing the behavior of the critical point {\bf GB3}. This saddle point corresponds to a de Sitter solution.}
    \label{fig:fig01}
\end{figure}
\end{widetext}

\noindent {\bf GB3}:  $(x_c,y_c,g_c)=(0,3,0)$,\; $\Omega_m=0$, \; $w_{eff}=-1$. This point satisfies the conditions for critical point, according to Eqs. (\ref{eq20})-(\ref{eq26}). This point is dominated by the scalar field potential and corresponds to de Sitter solution. According to the numerical analysis, as shown in the sample in Fig. 1, the point is a saddle point that attracts from some trajectories and repels trajectories that approach toward the attractor {\bf GB5}, as shown in fig. \ref{fig:fig01}.

\begin{widetext}
	
\begin{figure}[h]
    \centering
    \includegraphics [scale=0.25]{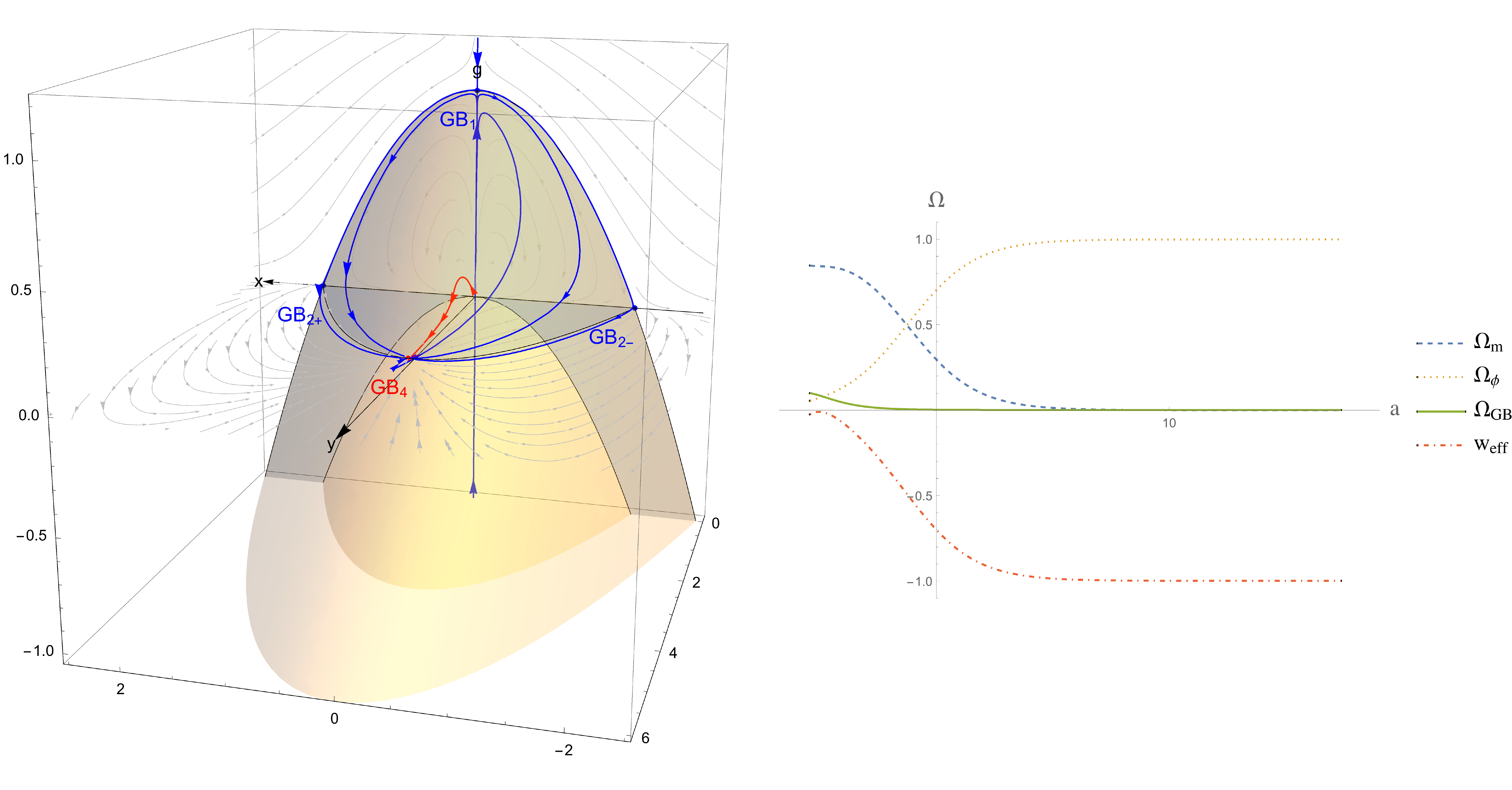}
    \caption{The phase portrait of the model for the parameters ($\beta=\frac{1}{16}$, $\lambda=\frac{1}{8}$, $w_m=0$). The graphic shows trajectories coming from different critical points that evolve towards the attractor {\bf GB4}. The shaded region corresponds to the region between the surfaces $\Omega_m=0$ and $\Omega_m=1$, i.e. delimits the physically allowed region that satisfies the restriction $0\le \Omega_m\le 1$. On the right side we plotted the evolution of the density parameters $\Omega_m, \Omega_{\phi}, \Omega_g$ and the EoS $w_{eff}$ along the trajectory (in red color) that goes from the matter dominated point $(0,0,0)$ to the scalar-field dominated fixed point {\bf GB4} .}
    \label{fig:fig02}
\end{figure}
\end{widetext}

\noindent {\bf GB4}:  $(x_c,y_c,g_c)=(\frac{\lambda}{2},3-\frac{\lambda^2}{4},0)$, with eigenvalues $[\frac{1}{2}(\beta-\lambda)\,\lambda,\frac{1}{4}(\lambda^2-12),\frac{1}{2}(\lambda^2-6\,w_m-6)]$. The density parameters and EoS are $\Omega_m=0$, \hspace{0.1cm} $\Omega_{\phi}$=1,\hspace{0.1cm} $w_{eff}=-1+\frac{\lambda^2}{6}$. This critical point is characterized by the scalar field dominance. At this point the potential is positive whenever  $\lambda<2\sqrt{3}$, which allows $-1<w_{eff}<1$. It is stable fixed point for $\lambda>\beta$ and $\lambda<\sqrt{6\,(1+w_m)}$ (whenever $0<w_m<1$). Saddle point for $\lambda<\beta, \sqrt{6(1+w_m)}$ or $\lambda>\beta$ and $\sqrt{6\,(1+w_m)}<\lambda<2\,\sqrt{3}$. Fig. \ref{fig:fig02} shows the phase portrait of the system with a set of parameters for which this fixed point is an stable attractor and all the trajectories diverging from the other critical points.

On the right side of Fig. \ref{fig:fig02} we show the evolution of $\Omega_m$, $\Omega_{\phi}$, $\Omega_g$ and $w_{eff}$ for $\lambda=1/8$, $\beta=1/16$, $w_m=0$, where the initial conditions have been adjusted to satisfy $\Omega_{m0}\approx 0.3$ and $w_{eff0}\approx -0.7$. The solution approaches the fixed point {\bf GB4} dominated by the scalar field.\\

\begin{widetext}

\begin{figure}[h]
    \centering
   \includegraphics [scale=0.25]{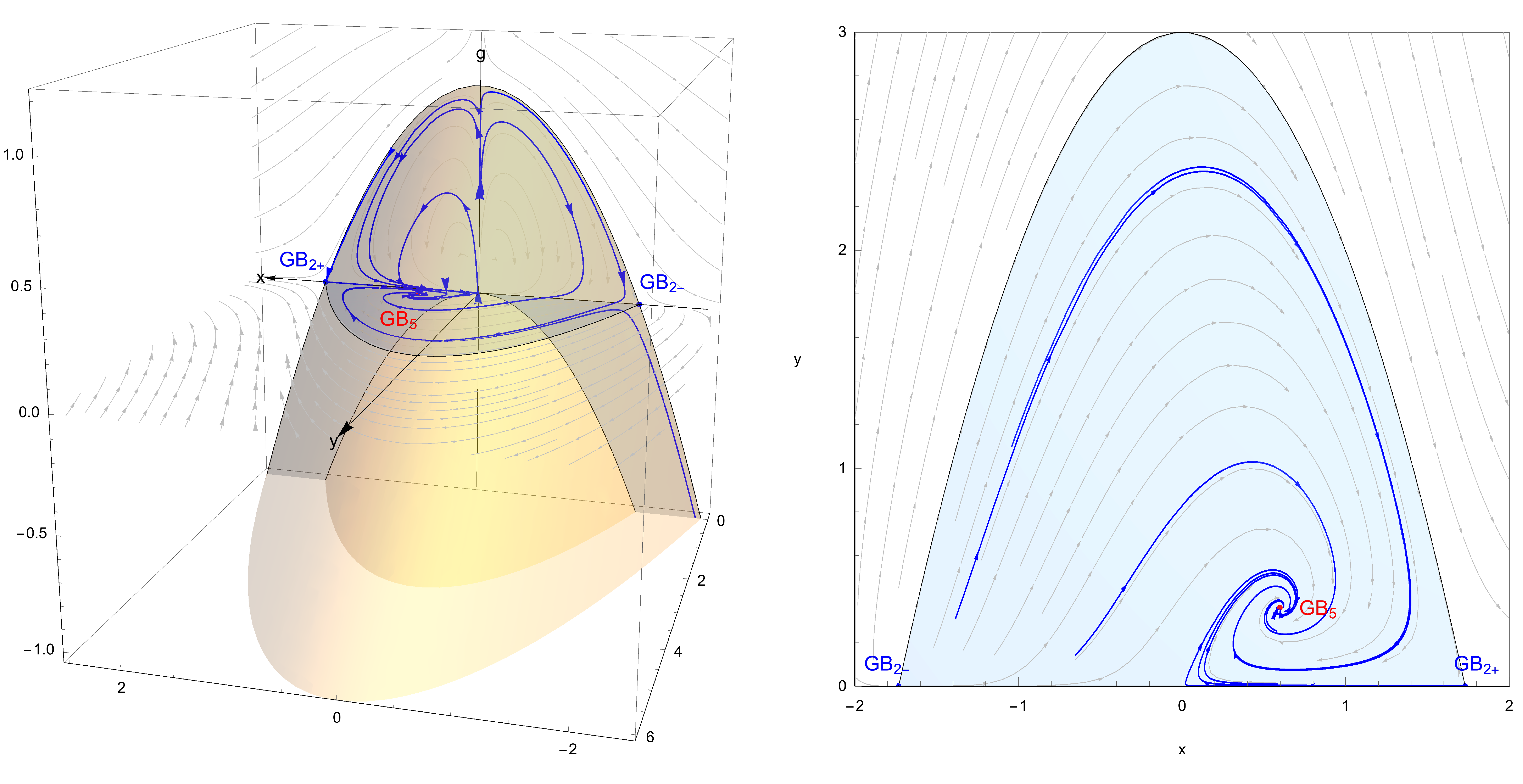}
    \caption{The phase portrait of the model for the set ($\beta=1$, $\lambda=5$, $w_m=0$) for which the point {\bf GB5} is a stable spiral. The graphic shows the projection on the $xy$ plane in which all the trajectories inside the region $0\leq\Omega_m\leq 1 $ tend to the point {\bf GB5}. The solution approaches the scaling solution {\bf GB5} with $\Omega_{\phi}=6/\lambda^2$ and $\Omega_m=1-6/\lambda^2$.}
    \label{fig:fig03}
\end{figure}
\end{widetext}

\noindent {\bf GB5}:  $(x_c,y_c,g_c)=(\frac{3(1+w_m)}{\lambda},\frac{9(1-w_m^2)}{\lambda^2},0)$, with eigenvalues $[\frac{3(1+w_m)(\beta-\lambda)}{\lambda},\frac{3(-1+w_m)}{4}-\frac{3\sqrt{\gamma_1}}{4\lambda},\frac{3(-1+w_m)}{4}+\frac{3\sqrt{\gamma_1}}{4\lambda}]$ (see below for 
$\gamma_1$). The main parameters are $\Omega_{\phi}=\frac{6(1+w_m)}{\lambda^2}$, \; $\Omega_m=1-\frac{6(1+w_m)}{\lambda^2}$,\; $w_{eff}=w_m$. Since $0\le\Omega_m\le 1$, then the critical point has physical meaning only in the case $\lambda\ge\sqrt{6(1+w_m)}$. 
This point corresponds to scaling solution and is a stable fixed point for $\beta<\lambda$ and $\sqrt{6(1+w_m)}<\lambda\le\sqrt{\frac{48}{7+9w_m}}(1+w_m)$ (provided that  $0\le w_m<1$).
It is a stable spiral for $0<w_m<1$,  $\beta<\lambda$ and $\lambda>\sqrt{\frac{48}{7+9w_m}}(1+w_m)$. 
Therefore, it is a saddle point  when $\beta>\lambda$. In Fig. \ref{fig:fig03} we show the phase portrait of the model with a set of parameters for which this point is an attractor (stable spiral).
\begin{equation*}\label{gama1}
\gamma_1=\left(w_m-1\right)\left(\lambda^2\left(9w_m+7\right)-48\left(w_m+1\right)^2\right)
\end{equation*}
\noindent {\bf GB6}: $(x_c,y_c,g_c)=(\frac{3(1+w_m)}{\beta},0,\frac{18(1-w_m)(1+w_m)^2}{\beta^2(1+3w_m)})$, with eigenvalues\\
$[\frac{3\,(-1+w_m)}{4}-\frac{3\,\sqrt{\gamma_2}}{4},\frac{3\,(-1+w_m)}{4}+\frac{3\,\sqrt{\gamma_2}}{4},\frac{3\,(1+w_m)(\beta-\lambda)}{\beta}]$ (see below for $\gamma_2$). The main parameters are $\Omega_{\phi}=\frac{3(1+w_m)^2}{\beta^2}$, $\Omega_m=1-\frac{3(1+w_m)^2(7-3w_m)}{(1+3w_m)\beta^2}$, \hspace{0.2cm} $w_{eff}=w_m$. This is a scaling solution without potential, and from the coordinates of this point follows that it exists whenever $\beta^2>\frac{3(1+w_m)^2(7-3w_m)}{1+3w_m}$ (assuming $0\le w_m\le 1$). From the expressions for the eigenvalues we have found that is not possible to satisfy the conditions of stability  with $0\le\Omega_m\le 1$. The point is saddle point for  $\beta>\sqrt{\frac{3(7+11w_m+w_m^2-3w_m^3)}{1+3w_m}}$ and $\lambda>\beta$. For the points {\bf GB5} and {\bf GB6} the matter and scalar field contribute with non-vanishing fraction to the total energy density.
\begin{widetext}
\begin{equation*}\label{gama2}
\gamma_2=\frac{(1 - w_m) (3 (1 + w_m)^2 (-53 + w_m (-233 + 15 w_m (-1 + 3 w_m))) +(9 + 7 w_m) (\beta + 3 w_m \beta)^2)}{9 (-1 + w_m) (1 + w_m)^2 (-1 + 9 w_m) + (\beta + 3 w_m \beta)^2}
\end{equation*}
\end{widetext}

\noindent {\bf GB7}: This point is dominated by the scalar field (the kinetic term and the GB coupling), giving $\Omega_m=0$. The expression for this point is too large and is given by the Eq. (\ref{eq100}) in the appendix, but we can consider an special case by fixing the parameter $\beta$. For $\beta=10\sqrt{5/27}$ this critical point takes the value $(x,y,g)=(\frac{3}{5},0,\frac{4}{5})$, leading to $\Omega_m=0$, $\Omega_{\phi}=1/5$ and $w_{eff}=1/9$, which is stable whenever $w_m>1/9$ and $\lambda>10\sqrt{5/27}$, and for $w_m=0$ is a saddle point.\\
\noindent On the other hand, in the limit of large $\beta$ ($\beta\rightarrow \infty$) it is obtained that $\Omega_{\phi}\rightarrow 0$ and the solution becomes dominated by the GB coupling, $\Omega_g\rightarrow 1$. At large $\beta$, the critical point takes the value $(0,0,1)$ with eigenvalues $[2,2,-1-3w_m]$, and the effective EoS tends to $w_{eff}\rightarrow -1/3$. Therefore the point becomes a saddle point describing the limit between decelerated and accelerated expansion. \\

\noindent Resuming, the analysis of the critical points shows that the model contains two relevant scaling solutions, one of them given by the point ({\bf GB5}) associated with the scalar potential, which  can be stable fixed point, stable spiral or saddle point, depending on the restrictions on $\lambda$ and $\beta$. This scaling solution can be used to describe the cosmological epoch in which the energy density of the scalar field behaves proportionally to that of the background fluid in either a radiation or matter domination era. The second solution ({\bf GB6}) due to the GB coupling, is a saddle point in the physical region of the density parameters, and therefore is not viable as scaling attractor solution. The point  ({\bf GB7}) is also related to the GB coupling, and becomes a saddle point in the limit of large $\beta$, leading to the solution in which the effective equation of state is $w_{eff}=-1/3$, describing an universe in the limit between decelerated and accelerated expansion. For the critical point ({\bf GB4}) the evolution is dominated by the scalar field and gives a stable solution with accelerated expansion provided that $\lambda^2<4$. This restriction to the values of $\lambda$ is interesting for dark energy solutions, but does not apply to matter or radiation scaling solutions where we need larger values of $\lambda$ in order to satisfy the  primordial nucleosynthesis constraint \cite{Ferreira:1997au}. It is worth noting that the point {\bf GB4} coincides with the point {\bf GB3} in the limit $\lambda\rightarrow 0$, but if we neglect the coupling term, then the dimension of the system reduces to two and the eigenvalues take the values $[-3,-3(1+w_m)]$. Thus the point {\bf GB3} (or {\bf GB4} in the limit $\lambda\rightarrow 0$) in absence of coupling becomes a de Sitter attractor.
 

\subsection{\label{sec:KC}The kinetic coupling (limit $g=0$)}
This limit corresponds to the scalar field with non-minimal kinetic coupling, which has been considered in \cite{Granda:2009fh}, \cite{Granda:2010hb}. Setting $g=0$ and $\beta=0$ in the system (\ref{eq20})-(\ref{eq26}) and solving the resulting system for $x'=0, y'=0, k'=0$, we find the following critical points:\\

\noindent {\bf K1}: $(x_c,y_c,k_c)=(0,0,1)$, with eigenvalues $[3,\frac{3}{2}, -3 w_m]$. The main parameters take the values: $\Omega_{\phi}=\Omega_m=0$, $\Omega_k=1$ and $w_{eff}=0$. This is unstable fixed point and corresponds to an evolution dominated by the kinetic coupling term. This kinetic-dominated solution is expected to be relevant at early times.\\

\noindent {\bf K$2_{\pm}$}: $(x_c,y_c,k_c)=(\pm \sqrt{3},0,0)$, with eigenvalues\\ $[-3 (-1 + w_m), - (6 \mp \sqrt{3}\alpha), 6 \mp \sqrt{3} \lambda]$. The density parameters take the values $\Omega_m=0$, \hspace{0.2cm} $\Omega_{\phi}=1$, and the EoS $w_{eff}=1$  corresponds to ``stiff matter'', evolution governed by the kinetic term. The solution with $\dot{\phi}>0$ is a saddle point whenever $w_m<1$, $\alpha<2\sqrt{3}$, $\lambda>2\sqrt{3}$ or for $\lambda,\alpha<2\sqrt{3}$ or $\lambda,\alpha> 2\sqrt{3}$.\\

\noindent {\bf K3}:\,$(x_c,y_c,k_c)$=($\frac{3(1+w_m)}{\alpha},0,\frac{9(1+w_m-w_m^2-w_m^3)}{2w_m\alpha^2}$), with eigenvalues\\ $[\frac{3(-1+w_m)}{4}-\frac{3\sqrt{\gamma_3}}{4},\frac{3(-1+w_m)}{4}+\frac{3\sqrt{\gamma_3}}{4},\frac{3(1+w_m)(\alpha-\lambda)}{\alpha}]$ (see below for $\gamma_3$). This critical point is a new scaling solution dominated by the scalar field with the kinetic coupling, and is a finite fixed point for $w_m\ne 0$. The effective equation of state is $w_{eff}=w_m$, and $\Omega_{\phi}=\frac{3(1+w_m)^2}{\alpha^2}$,  $\Omega_m=\frac{3w_m^3+2\alpha^2 w_m-3w_m^2-15w_m-9}{2w_m\alpha^2}$. This point exists whenever $0\le\Omega_m\le 1$, which leads to the restriction $\alpha^2>\frac{9+15w_m+3w_m^2-3w_m^3}{2w_m}$ ($0<w_m<1$). Nevertheless, in this case it is not possible to simultaneously satisfy the stability conditions for the critical point and the restriction on $\Omega_m$.  The point is saddle for $\lambda>\alpha$ and $\alpha>\sqrt{\frac{9+15w_m+3w_m^2-3w_m^3}{2w_m}}$, which gives two negative eigenvalues.
\begin{widetext}

\begin{figure}[h]
    \centering
   \includegraphics [scale=0.25]{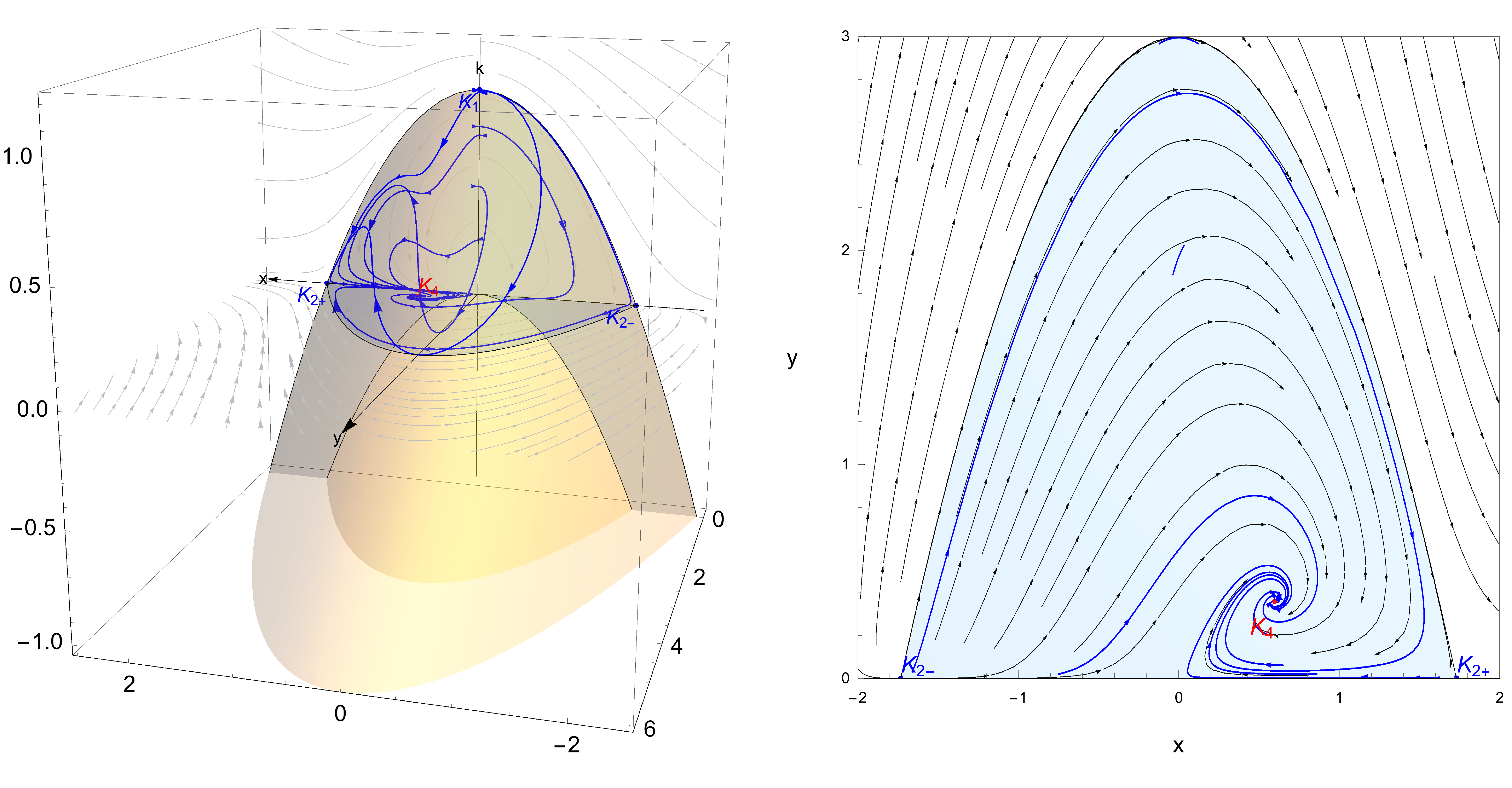}
    \caption{The phase portrait of the model for the parameters ($\alpha=2$, $\lambda=5$, $w_m=0$) showing the scaling attractor solution  {\bf K4}. The graphic shows trajectories departing from the point {\bf K1} and the saddle points ${\bf K2_{+}}$,  ${\bf K2_{-}}$ to finally end at the fixed point {\bf K4}, which is a stable spiral. The behavior along the $k$-axis reflects the fact that when $x=y=0$  (provided that $w_m=0$), the equation for $k'=0$ is automatically satisfied for any $k$.}
    \label{fig:fig04}
\end{figure}
\end{widetext}
\begin{equation*}
\begin{aligned}
\gamma_3=&B\big(1-w_m\big)\big(9-24w_m^3+33w_m^4+2w_m(\alpha^2-84)\\ &+6w_m^2(5\alpha^2-39)\big)
\end{aligned}
\end{equation*}
whrere $B=1/\big(9-9w_m^2+15w_m^3+w_m(2\alpha^2-15)\big)$\\

\noindent {\bf K4}: $(x_c,y_c,k_c)=(\frac{3(1+w_m)}{\lambda}, \frac{9(1-w_m^2)}{\lambda^2},0)$, with eigen\-values $[\frac{3(1+w_m)(\alpha-\lambda)}{\lambda},\frac{3(-1+w_m)}{4}-\frac{3\sqrt{\gamma_4}}{4\lambda},\frac{3(-1+w_m)}{4}+\frac{3\sqrt{\gamma_4}}{4\lambda}]$ (see below for $\gamma_4$). This critical point is a scaling solution provided by the scalar field with exponential potential. The main parameters are $w_{eff}=w_m$, \hspace{0.2cm} $\Omega_{\phi}=\frac{6(1+w)}{\lambda^2}$, \hspace{0.2cm} $\Omega_m=1-\Omega_{\phi}$. This point is physical ($0\le\Omega_m\le 1$) whenever $\lambda>\sqrt{6(1+w_m)}$.
The point is stable spiral for $\lambda>\frac{4\sqrt{3}(1+w_m)}{\sqrt{7+9w_m}}$ and $\alpha<\lambda$, provided  $0\le w_m<1$. It is stable fixed point when  $\sqrt{6(1+w_m)}<\lambda\le\frac{4\sqrt{3}(1+w_m)}{\sqrt{7+9w_m}}$ and $\alpha<\lambda$. When $\alpha>\lambda$ we have a saddle critical point. In Fig. 4 we show the phase portrait of the model where the point  {\bf K4} is an stable spiral.
Note that for the points {\bf K3} and {\bf K4} the matter and scalar field contribute both with a non-vanishing fraction to the total energy density.
$$\gamma_4=(-1+w_m)\left((7+9w_m)\lambda^2-48(1+w_m)^2\right)$$

\begin{widetext}

\begin{figure}[h]
    \centering
    \includegraphics[scale=0.25]{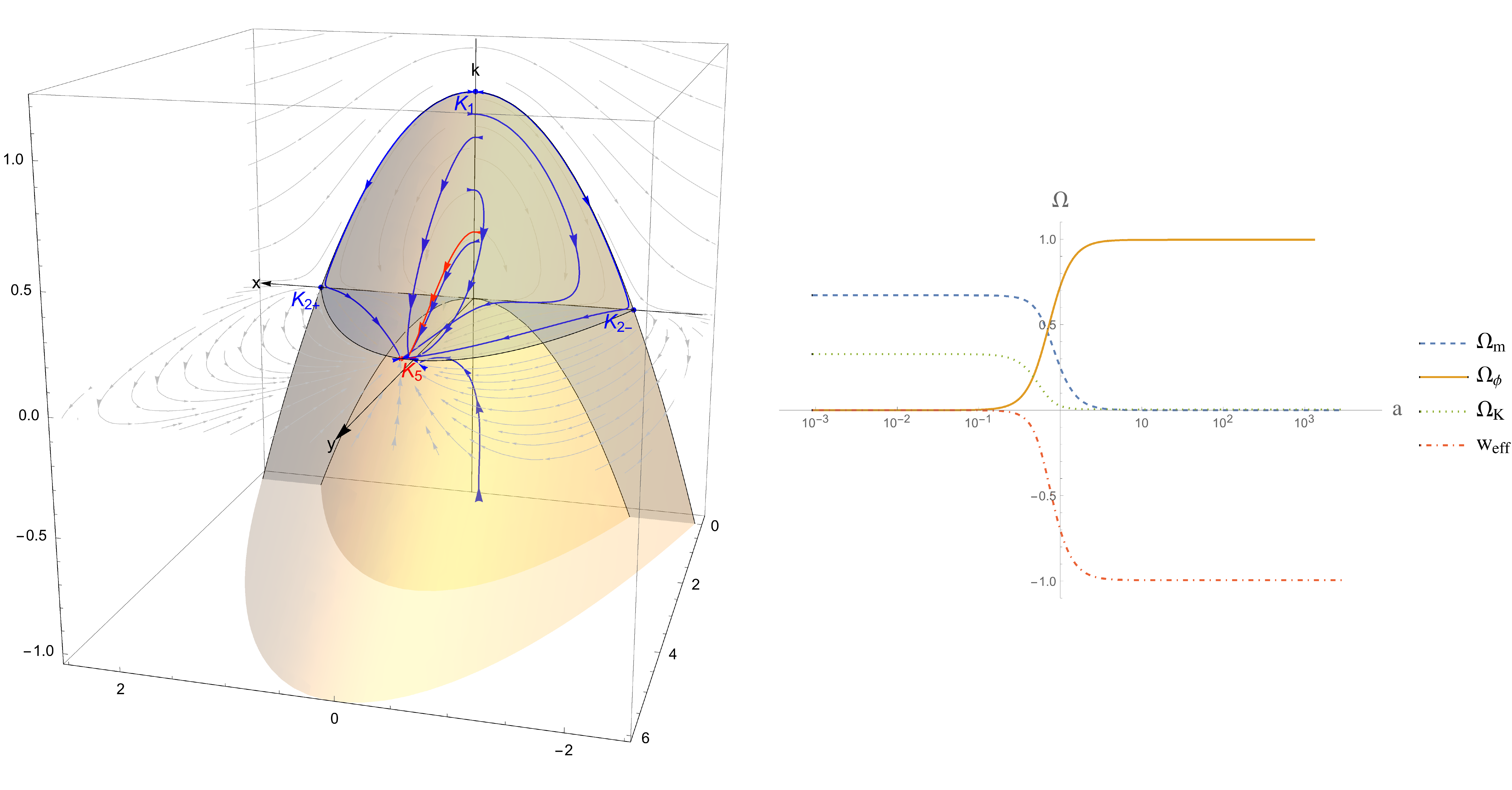}
    \caption{The phase portrait of the model for the set ($\alpha=\frac{1}{8}$, $\lambda=\frac{1}{4}$, $w_m=0$) showing trajectories  that converge to the scalar field dominated fixed point {\bf K5} which is a late-time  attractor  solution corresponding to accelerated expansion. The graphic on the right shows the evolution of of the density parameters $\Omega_m$, $\Omega_{\phi}$, $\Omega_g$ and the EoS $w_{eff}$ along the trajectory (in red color) that starts in a point that repels from the $k$-axis and ends in the scalar-field dominated fixed point {\bf K5}.}
    \label{fig:fig05}
\end{figure}
\end{widetext}

\noindent {\bf K5}: $(x_c,y_c,k_c)=(\frac{\lambda}{2}, \frac{1}{4}(12-\lambda^2),0)$, with eigenvalues $[-3+\frac{\lambda^2}{4},-3(1+w_m)+\frac{\lambda^2}{2},\frac{1}{2}\lambda(\alpha-\lambda)]$. This fixed point corresponds to scalar field domination, with $w_{eff}=-1+\frac{\lambda^2}{6}$, $\Omega_{\phi}=1$, $\Omega_m=0$. This point is stable attractor  for $\lambda<\sqrt{6(1+w_m)}$, whenever $0\le w_m \le 1$, and $\alpha<\lambda$. It is a saddle point if $\lambda>\sqrt{12}$ and $\alpha<\lambda$ (this case is not of interest because the potential becomes negative) or $\lambda<\sqrt{6(1+w_m)}$ and $\alpha>\lambda$. The phase portrait of the model, showing the attractor quality of the point {\bf K5} is shown in  Fig. \ref{fig:fig05}.
The initial conditions for the density parameters and the EoS $w_{eff}$ in Fig. \ref{fig:fig05} have been choosen such that at the present $\Omega_{m0}\approx 0.3$ and $w_{eff0}\approx -0.7$. Note that the standard quintessence scaling solution takes place when the kinetic coupling disappears.

\noindent {\bf K6}: this point is dominated by the scalar field (the kinetic term and the kinetic coupling). The expression for this point is too large and is given by the Eq. (\ref{eq101}) in the appendix, but we can consider an special case by fixing the parameter $\alpha$. For $\alpha=8/\sqrt{3}$ this fixed point takes the value $(x,y,k)=(\frac{\sqrt{3}}{2},0,\frac{3}{4})$, giving $\Omega_m=0$, $\Omega_{\phi}=1/4$, $w_{eff}=1/3$, and is stable whenever $w_m>1/3$ and $\lambda>8\sqrt{3}$. For $w_m=0$ is a saddle point. Therefore, this point is an stable attractor for the scalar field mimicking radiation dominated universe.\\
\noindent On the other hand, in the large $\alpha$ limit, the critical point takes the value $(0,0,1)$ with eigenvalues $[3,3/2,-3w_m]$.  At this limit the solution becomes dominated by the kinetic coupling and the point is saddle. The effective EoS tends to $w_{eff}\rightarrow 0$, indicating that the scalar field mimics the pressureless matter dominated universe. \\
According to the above results, the scalar filed with the kinetic coupling contains two relevant scaling solutions. The first one is the scaling solution {\bf K4}  connected with the scalar field potential  and behaves similar  to the solution {\bf GB5} obtained for the model with GB coupling. The point {\bf K3} is the new scaling solution dominated by the scalar field with the kinetic coupling. As in the case of the GB coupling, the point is not stable in the allowed region of the physical parameters, and behaves as a saddle point. The point {\bf K6} dominated by the free kinetic term and the kinetic coupling, which depends on the parameter $\alpha$, gives two interesting results: by setting $\alpha$ to the value $\alpha=8\sqrt{3}$ the fixed point is an attractor that mimics the radiation dominated universe, and in the limit of large $\alpha$ the point becomes a saddle point dominated by the kinetic coupling and mimics presureless matter dominated universe.  The point {\bf K5}, which is important for late time universe, presents a behavior similar to the point {\bf GB4} and in the absence of the kinetic coupling (in the limit $\lambda\rightarrow 0$) is a de Sitter attractor.
\subsection{\label{sec:GBKT}The Gauss-Bonnet and kinetic couplings}
The model presents 11 critical points, two of them lie on the plane $xy$, two on the plane $xk$, two on the plane $xg$, one on the plane $kg$, two on the $x$ axis, one along the $k$ axis and one along the $g$ axis. There are three fixed points, on the planes $xy$, $xk$ and $xg$, that present scaling behavior, and are stable attractors or behave as saddle points, depending on the regions (swept) by  the parameters $\alpha$, $\beta$, $\lambda$. There is also one point in the $xy$-plane that is a quintessence solution with the stability properties depending on the correlation between the parameters  $\alpha$, $\beta$, $\lambda$.\\

\noindent {\bf A}: ($x_c,y_c,k_c,g_c)=(0,0,1,0)$\hspace{1cm}with eigenvalues $[3, -3/2, 3/2, -3 w_m]$. \hspace{0.1cm}$\Omega_{\phi}=\Omega_m=0$, $\Omega_k=1$ and $w_{eff}=0$. This is unstable fixed point and corresponds to an evolution dominated by the kinetic coupling term. This kinetic-dominated solution is expected to be relevant at early times.\\

\noindent {\bf B}: $(x_c,y_c,k_c,g_c)=(0,0,0,1)$\hspace{1cm} with eigenvalues $[2, 2, 2, -1 - 3 w_m]$ . \hspace{0.1cm} $\Omega_m=0$, \hspace{0.1cm} $\Omega_{GB}=1$, \hspace{0.1cm} $w_{eff}=-1/3$, in this case the evolution is dominated by the Gauss-Bonnet term, and $w_{eff}$ corresponds to the divide between decelerated and accelerated expansion.  This unstable fixed point is saddle for $w_m>-1/3$.\\

\noindent {\bf C}: $(x_c,y_c,k_c,g_c)=(0,3(1+k),k,-2k)$\hspace{1cm} with eigenvalues $[0, 0, -3, -3(1+w_m)]$. This critical point is dominated by the scalar potential and the coupling terms, but  $k$ is arbitrary, which indicates that any point on the $k$-axis is a critical point in this case. The fact that the eigenvalues do not depend on $k$ suggests that all the points on the $k$-axis have the same stability properties, but the presence of two zero eigenvalues indicates that in order to evaluate the stability of these points, we have to consider expansions (around the fixed point) beyond the linear order in the dynamical equations. The density parameters take the values $\Omega_m=0$, $\Omega_{\phi}=1+k$, $\Omega_g=-2k$, $\Omega_k=k$, where the fact that $\Omega_{\phi}>1$ is compensated with the opposite sign of the coupled term $\Omega_g$. In fact we can consider the sum $\Omega_{\phi}+\Omega_k+\Omega_g$ as the effective energy density parameter of the scalar field, due to the interaction terms. The EoS of this fixed point corresponds to a de Sitter solution $w_{eff}=-1$, which is marginally stable, since it has two negative and two zero eigenvalues. To analyze the stability we fix a constant value for $k$ in order to have one critical point and apply the central manifold technique. The central manifold analysis imposes restrictions on the constant $k$ and shows that the point ${\bf C}$ is a saddle point (see appendix B for details). \\ 

\noindent {\bf D}: $(x_c,y_c,k_c,g_c)=(\pm \sqrt{3},0,0,0)$, with eigenvalues $[3(1-w_m),-6\pm\sqrt{3}\alpha,-6\pm\sqrt{3}\beta,6\mp\sqrt{3}\lambda]$. The main parameters take the values \hspace{0.1cm} $\Omega_m=0$, \hspace{0.2cm} $\Omega_{\phi}=1$, \hspace{0.2cm}$w_{eff}=1$, ``stiff matter'', evolution governed by the kinetic term. The solution with $\dot{\phi}>0$ is an unstable fixed point when $w_m<1$, $\alpha, \beta>2\sqrt{3}$ and $\lambda<\sqrt{2}$, and is a saddle point  whenever $w_m<1$ and at least one of the above inequalities is unsatisfied.\\

\begin{widetext}

\begin{figure}[h]
    \centering
    \includegraphics[scale=0.40]{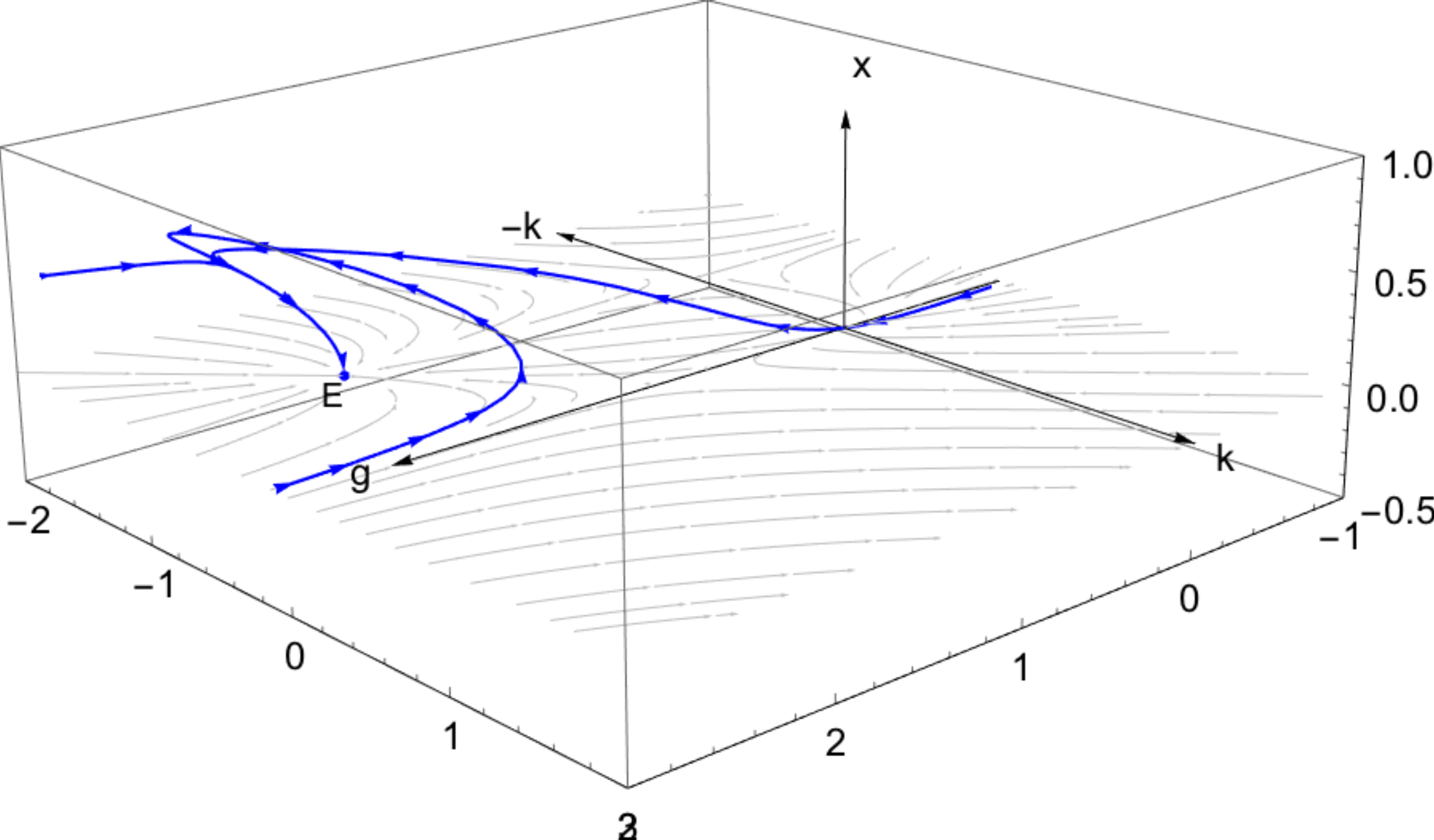}
    \caption{The phase portrait of the model for the set ($\beta=1/8$, $\alpha=1/4$, $w_m=0$) showing trajectories  that converge to the scalar field dominated attractor point {\bf E} in the $(x,k,g)$ coordinates. In this graphic the paths, plotted for different initial conditions, converge to fixed point {\bf E}.}   
    \label{fig:fig06}
\end{figure}
\end{widetext}

\noindent {\bf E}: $(x_c,y_c,k_c,g_c)=(0,0,-1,2)$, \hspace{1cm}with eigenvalues $[-3, 0, 0, -3 (1 + w_m)]$. The main parameters are $\Omega_m=\Omega_{\phi}=0$, \hspace{0.1cm}$w_{eff}=-1$. In this critical point the evolution is dominated by the kinetic and Gauss-Bonnet couplings and corresponds to a de Sitter solution. Due to the zero eigenvalues the analysis of stability of this point is complicated (for $w_m>-1$, the point is marginally stable), and demands further analysis. The analysis of the stability around this point can be simplified if we consider the system without potential term. In this case the dimension of the system reduces to three (i.e. $(x_c,k_c,g_c)$=($0,-1,2$)) and the eigenvalues will be $[-3, 0, -3 (1 + w_m)]$. The numerical analysis demonstrates that the point is a stable fixed point as shown in Fig. \ref{fig:fig06}. Thus, in absence of the scalar potential, the non-minimal kinetic and GB terms drive the system towards a de Sitter attractor.\\

\noindent {\bf F}:$(x_c,y_c,k_c,g_c)=(\frac{3(1+w_m)}{\alpha},0,\frac{9(1+w_m-w^2-w^3)}{2\alpha^2\,w_m},0$), with eigenvalues $[\frac{3(-1+w_m)}{4}-\frac{3\sqrt{\gamma_3}}{4},\frac{3(-1+w_m)}{4}+\frac{3\sqrt{\gamma_3}}{4},-\frac{3(1+w_m)(\alpha-\beta)}{\alpha},\frac{3(1+w_m)(\alpha-\lambda)}{\alpha}]$. The density parameters and the EoS are $\Omega_{\phi}=\frac{3(1+w_m)^2}{\alpha^2}$, $\Omega_m=\frac{3w_m^3-3w_m^2+2\alpha^2,w_m-15w_m-9}{2\alpha^2 w_m}$, \hspace{0.2cm}$w_{eff}=w_m$.
This critical point is a scaling solution dominated by the kinetic terms and is finite for $w_m\ne 0$. From $0\le\Omega_m\le1$ follows the restriction $\alpha^2>\frac{9+15w_m+3w_m^2-3w_m^3}{2w_m}$ (provided $0<w_m<1$). It is not possible to satisfy the stability conditions for $0\le \Omega_m\le 1$. This point is saddle whenever $\alpha>\sqrt{\frac{9+15w_m+3w_m^2-3w_m^3}{2w_m}}$ and $\beta<\alpha$ or $\lambda>\alpha$.\\

\noindent {\bf G}: $(x_c,y_c,k_c,g_c)$=($\frac{3(1+w_m)}{\beta},0,0,\frac{18(1-w_m)(1+w_m)^2}{\beta^2(1+3w_m)}$), with eigenvalues $[\frac{3(1+w_m)(\alpha-\beta)}{\beta},\frac{3(-1+w_m)}{4}-\frac{3\sqrt{\gamma_2}}{4}, \frac{3(-1+w_m)}{4}+\frac{3\sqrt{\gamma_2}}{4},\frac{3(1+w_m)(\beta-\lambda)}{\beta}]$. The density parameters and the EoS take the values $\Omega_{\phi}=\frac{3(1+w_m)^2}{\beta^2}$, $\Omega_m=\frac{\beta^2(1+3w_m)-33w_m-3w_m^2+9w_m^3-21}{\beta^2(1+3w_m)}$\hspace{0.2cm}$w_{eff}=w_m$. The condition $0\le\Omega_m\le 1$ for  $0\le w_m<1$ is satisfied if $\beta^2>\frac{21+33w_m+3w_m^2-9w_m^3}{1+3w_m}$. As in the point {\bf GB6}, in this case the point is unstable under the condition $0\le\Omega_m\le 1$. 
This scaling solution is a saddle fixed point for  $\beta>\sqrt{\frac{21+33w_m+3w_m^2-9w_m^3}{1+3w_m}}$ and $\alpha<\beta$ or $\beta<\lambda$.\\

\noindent {\bf H}: $(x_c,y_c,k_c,g_c)$=($\frac{3(1+w_m)}{\lambda},\frac{9(1-w^2)}{\lambda^2},0,0$), with eigen\-values $[\frac{3(1+w_m)(\alpha-\lambda)}{\lambda},\frac{3(1+w_m)(\beta-\lambda)}{\lambda},\frac{3(-1+w_m)}{4}-\frac{3\sqrt{\gamma_4}}{4\lambda},\frac{3(-1+w_m)}{4}+\frac{3\sqrt{\gamma_4}}{4\lambda}]$. The density parameters and the EoS become $\Omega_{\phi}=\frac{6(1+w_m)}{\lambda^2}$, $\Omega_m=\frac{\lambda^2-6(1+w_m)}{\lambda^2}$,\hspace{0.2cm}$w_{eff}=w_m$. This is a scaling solution with $\Omega_m\ne 0$, defined for $\lambda\ge\sqrt{6(1+w_m)}$, and becomes com\-ple\-te\-ly do\-mi\-na\-ted by the scalar field when $\lambda=\sqrt{6(1+w_m)}$. The point is an stable attractor when $\alpha, \beta<\lambda$ and $\sqrt{6(1+w_m)}<\lambda\le\sqrt{\frac{48}{7+9w_m}}(1+w_m)$ for $0\le w_m<1$. It is stable spiral when $\lambda>\sqrt{\frac{48}{7+9w_m}}(1+w_m)$ and is a saddle fixed point if $\alpha>\lambda$ or $\beta>\lambda$. The phase portrait of the system showing the point {\bf H} as an attractor is illustrated in Fig. 7.\\

\begin{widetext}

\begin{figure}[h]
    \centering
    \includegraphics[scale=0.25]{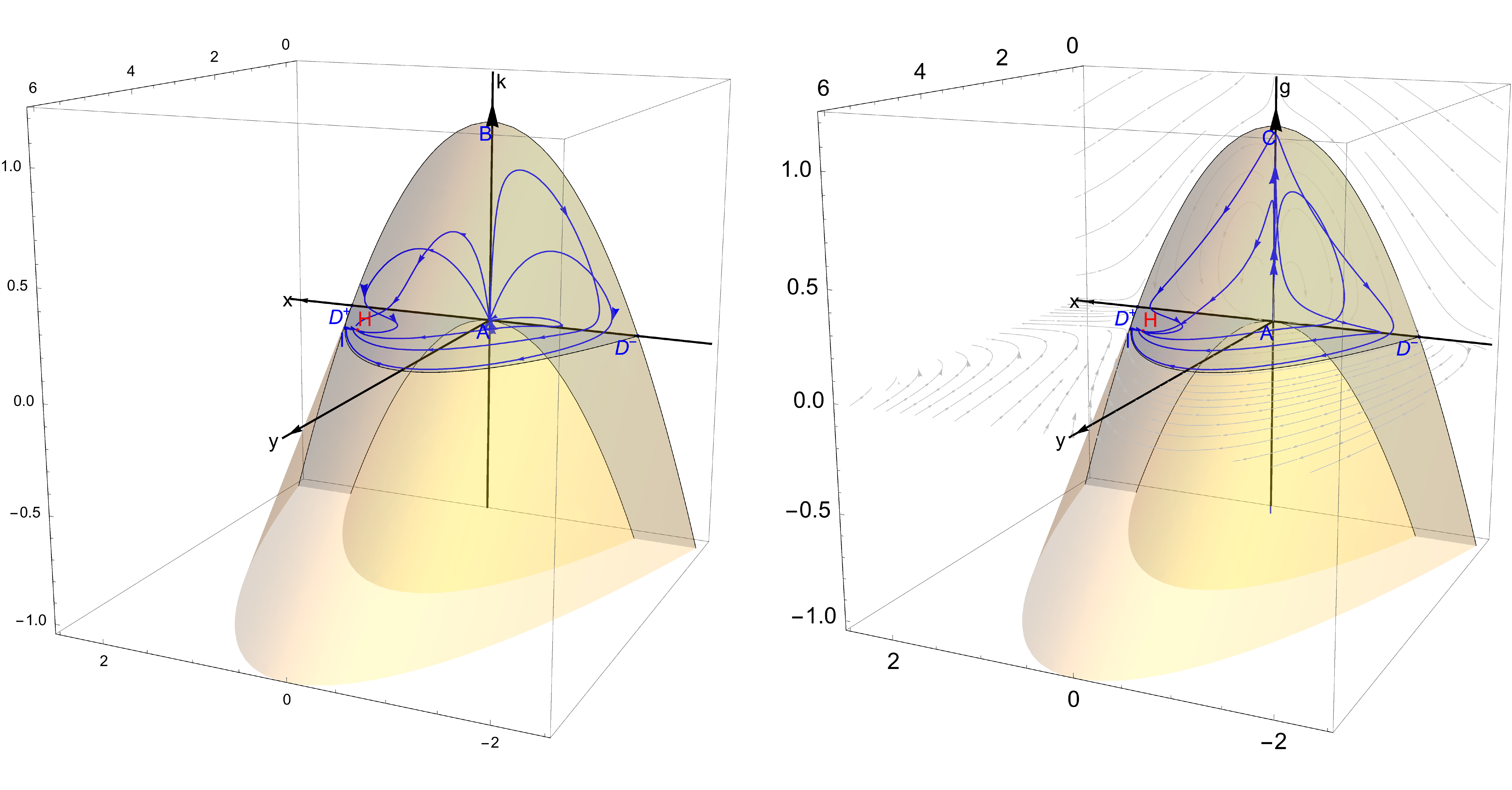}
    \caption{The phase portrait of the model for the set ($\beta=68/39$, $\lambda=4\sqrt{3/7}$, $\alpha=2$, $w_m=0$) showing trajectories  that converge to the scalar field dominated attractor point {\bf H} in the $(x,y,k)$ coordinates. The graphic on the right shows the phase portrait in the $(x,y,g)$ coordinates.}
    \label{fig:fig07}
\end{figure}
\end{widetext}

\begin{widetext}

\begin{figure}[h]
    \centering
    \includegraphics[scale=0.25]{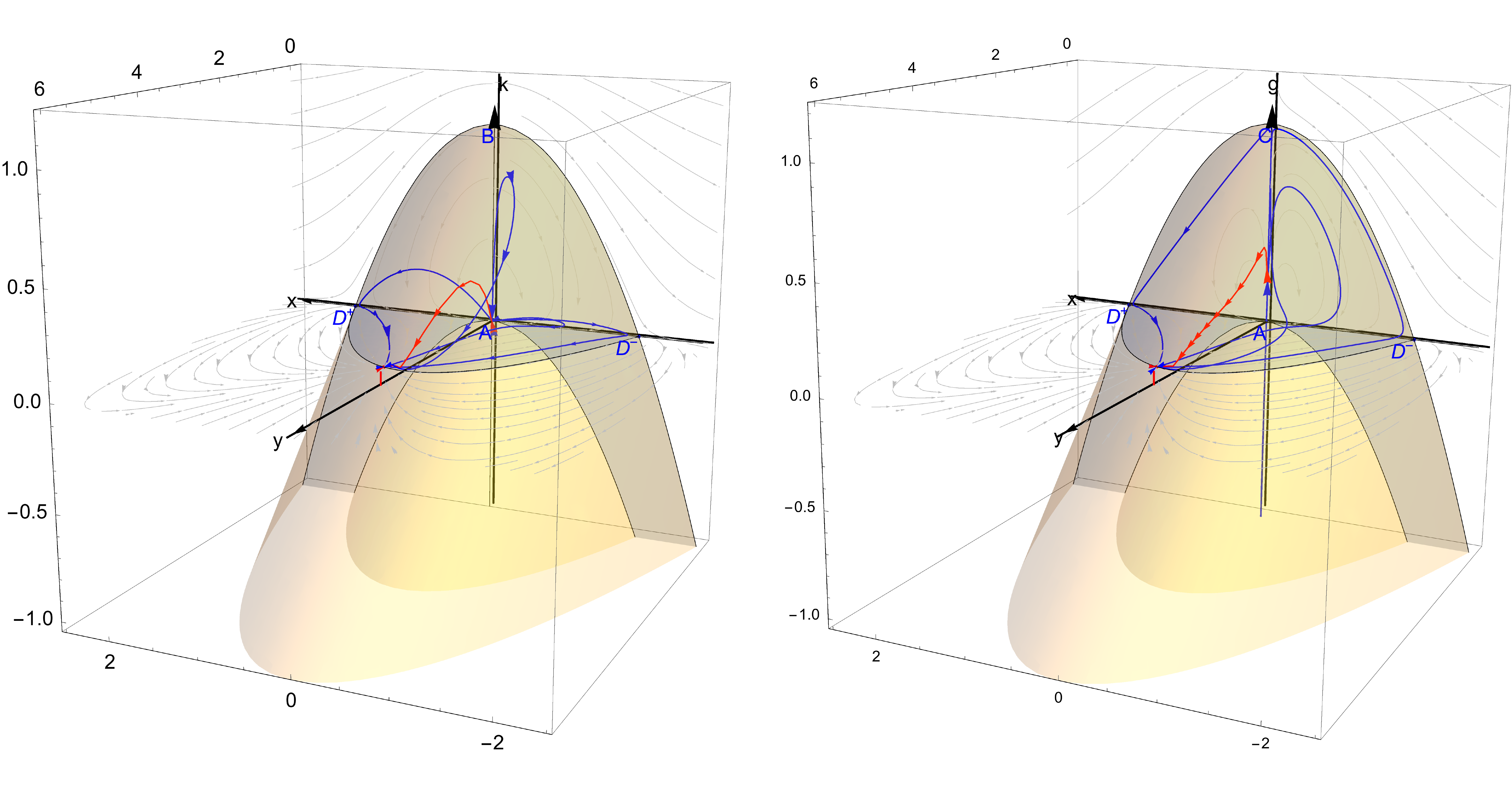}
    \caption{The phase portrait of the model for the set of parameters ($\beta=1/8$, $\lambda=1/2$, $\alpha=1/4$, $w_m=0$) showing trajectories  that converge to the scalar field dominated fixed point {\bf I}, which is a late-time attractor  solution corresponding to accelerated expansion. The graphic on the right shows the phase portrait in the $(x,y,g)$ coordinates.}
    \label{fig:fig08}
\end{figure}
\end{widetext}

\noindent {\bf I}: $(x_c,y_c,k_c,g_c)$=($\frac{\lambda}{2},3-\frac{\lambda^2}{4},0,0$), with eigenvalues $[\frac{1}{2}(\beta-\lambda)\lambda,-3+\frac{1}{4}\lambda^2,-3(1+w_m)+\frac{1}{2}\lambda^2,\frac{1}{2}(\alpha-\lambda)\lambda]$. The main parameters are $\Omega_{\phi}=1$, $\Omega_m=0$ and  $w_{eff}=-1+\frac{\lambda^2}{6}$. The positivity of the potential requires that $\lambda<2\sqrt{3}$. This critical point dominated by the uncoupled part of the scalar field leads to dark energy solution, and is stable when $\beta<\lambda$, $\alpha<\lambda$ and $\lambda<\sqrt{6(1+w_m)}$ (for $0\le w_m\le 1$). It is a saddle point when $\alpha>\lambda$ or $\beta>\lambda$. Note that if $\lambda^2=6(1+w_m)$, the $w_{eff}=w_m$ and the solution becomes scaling, but one of the eigenvectors becomes zero. This means that the point is marginally stable in this case. In Fig. \ref{fig:fig08}  we plot the phase portrait of the system in the $(x,y,k)$ coordinates, showing the attractor character of the point {\bf I}. Note that in absence of both couplings the dimension of the system reduces to two, with eigenvalues $[-3+\frac{1}{4}\lambda^2,-3(1+w_m)+\frac{1}{2}\lambda^2]$, and in the limit $\lambda\rightarrow 0$ the point becomes a de Sitter attractor.

\begin{figure}[h]
    \centering
    \includegraphics[scale=0.225]{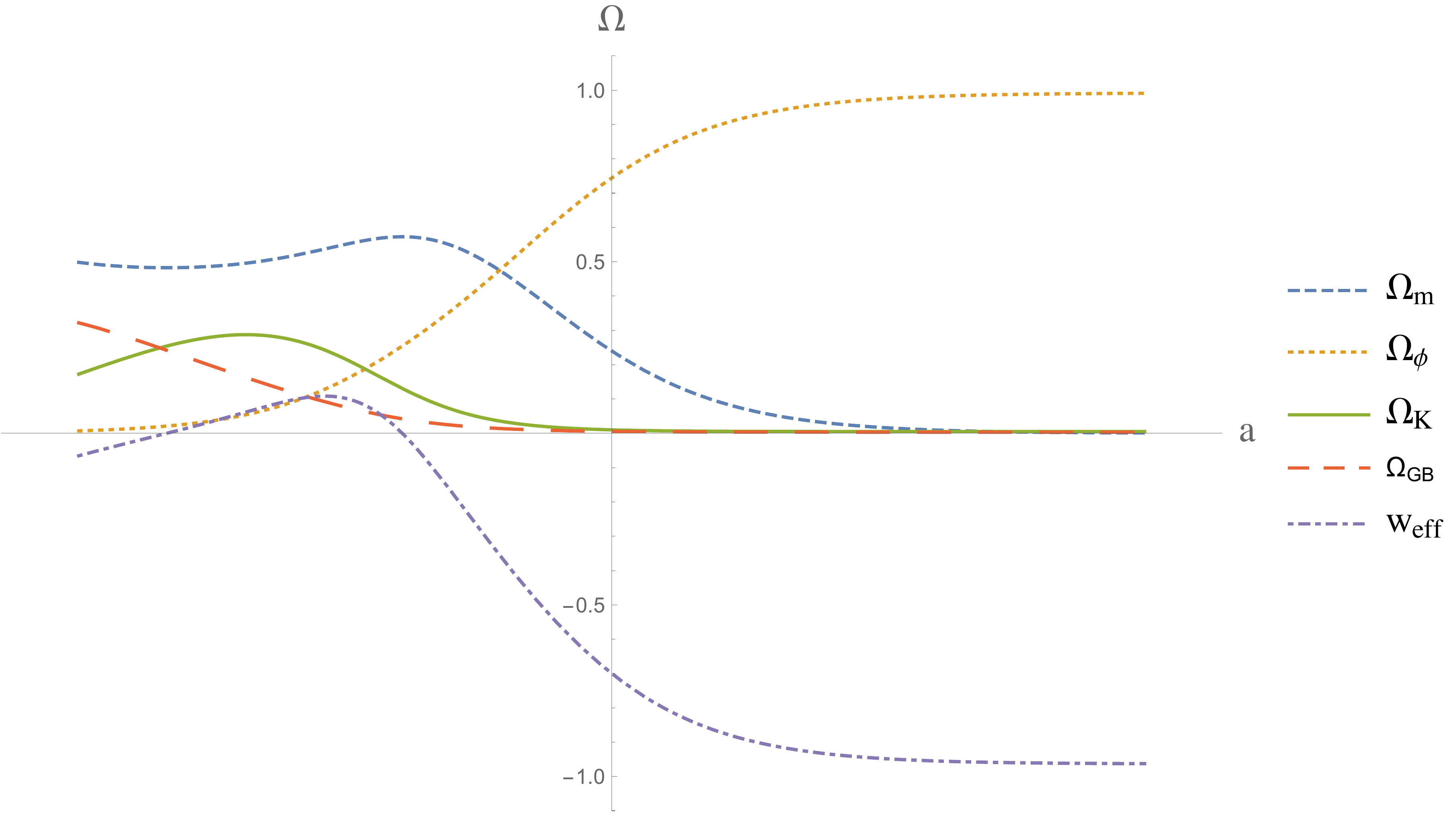}
    \caption{The evolution of of the density parameters $\Omega_m$, $\Omega_{\phi}$, $\Omega_k$, $\Omega_g$ and the EoS $w_{eff}$ along the trajectory (in red color) that pass close to the origen (the red curve in Fig. 8) and ends in the scalar-field dominated fixed point {\bf I}.}
    \label{fig:fig09}
\end{figure}

\noindent In Fig.\ref{fig:fig09} the graphic shows the evolution of the density parameters $\Omega_m, \Omega_{\phi}, \Omega_k, \Omega_m$ and the effective EoS $w_{eff}$ for the trajectories (red lines) plotted in Figs. 8. The initial conditions for the density parameters and $w_{eff}$ have been choosen such that at the present $\Omega_{m0}\approx 0.3$ and $w_{eff0}\approx -0.7$.\\

\noindent {\bf J}: There are two additional fixed points that have complicated dependence on $\alpha$ ({\bf J$_{\alpha}$}), where only the free kinetic and the coupled kinetic terms survive, and on $\beta$ ({\bf J$_{\beta}$}), where the free kinetic and coupled GB terms survive, given by Eqs. (\ref{eq100})  and (\ref{eq101}) respectively but don't lead to interesting late-time evolutionary scenarios. Nevertheless, the expressions for these critical points substantially simplify for $\alpha=8/\sqrt{3}$, $(\sqrt{3}/2,0,3/4,0)$, and $\beta=10\sqrt{5/27}$, $(\sqrt{3/5},0,0,4/5)$, as was shown for {\bf GB7} and {\bf K6}.
\noindent The analysis of the autonomous system with both couplings shows three scaling solutions, related in the different cases with the kinetic coupling ({\bf F}), the GB coupling ({\bf G}) and with the scalar potential ({\bf H}). The scaling solutions related with the couplings are saddle points in the physical region of parameters and follow the same behavior of the points {\bf K3} and  {\bf GB6}) for the kinetic and GB coupling respectively. The scaling solution dominated by the scalar potential can be saddle point, stable or stable spiral, depending on the values of $\lambda$. Particularly, for the stable spiral the values of $\lambda$ can satisfy the restriction imposed by primordial nucleosintesis. The scalar field dominated fixed point {\bf I} follows the same behavior described by the points {\bf GB4} and {\bf K5}. 
There is a new critical point that combines the effect of GB and kinetic couplings given by {\bf E }. In absence of potential the dynamical system reduces to three dimensions and the eigenvalues of the point {\bf E } reduce to $[-3,0,-3(1+w_m)]$, with two negative and one zero values. The results of the numerical analysis presented in Fig. 6 show that this point is a de Sitter attractor.\\

\noindent For the scaling solution {\bf K4} (or {\bf GB5}) we can analyze qualitatively the behavior of the non-minimally coupled kinetic term compared to that of the other terms in the model. In general for the scaling solution we have $H^2\sim a^{-3(1+w_m)}$, and from the equation for the critical point $x=3(1+w_m)/\lambda$, follows that (using the eq. (\ref{eq15}) for $x$) $\phi\sim (1+w_m)\ln a$. The energy density associated with the kinetic coupled term becomes $\rho_k=9\kappa^2 H^2 F_1\dot{\phi}^2=9\kappa^2H^4\phi'^2F_1\sim a^{-6(1+w_m)} e^{\frac{3\alpha}{\lambda}(1+w_m)\ln a}\sim a^{-3(1+w_m)(2-\frac{\alpha}{\lambda})}$. From this expression is clear that the contribution to the energy density coming from the kinetic coupling dilutes faster than the matter energy density whenever $\alpha<\lambda$, which is also a necessary condition for the stability of the scaling solution {\bf K4}. The GB contribution to the energy density for the same scaling solution behaves similarly as $\rho_{GB}\sim a^{-3(1+w_m)(2-\frac{\beta}{\lambda})}$, and also dilutes faster than the matter energy density in the case $\beta<\lambda$, which is a necessary stability condition for {\bf GB5}. This scaling solution also takes place when both couplings are present and corresponds to the point  {\bf G} (for the exponential couplings and potential as given in eq. (\ref{eq14})). In this case, under the conditions $\alpha<\lambda$ and $\beta<\lambda$, the kinetic and GB couplings become subdominant with respect to the uncoupled part of the scalar field at the present epoch. 

\begin{widetext}

\begin{table}[h]
    \centering
    \caption{ Summary of  eigenvalues}
\resizebox{\textwidth}{!}{ 
\begingroup
\squeezetable
   \renewcommand{\arraystretch}{2.5}
\begin{tabular}{lccccccccc}\hline\hline
 &
$x$&
$y$&
$k$&
$g$&
$w_{eff}$&
$\Omega_{m}$&
$\Omega_{\phi}$&
$\Omega_{k}$&
$\Omega_{g}$\\\hline
{\bf A}&$0$&$0$&$1$&$0$&$0$&$1$&$0$&$1$&$0$\\\hline
{\bf B}&$0$&$0$&$0$&$1$&$-1/3$&$0$&$0$&$0$&$1$\\\hline
{\bf C}&$0$&$3(1+k)$&$k$&$-2k$&$-1$&$0$&$1+k$&$k$&$-2kk$\\\hline
{\bf D$_\pm$}&$\pm \sqrt{3}$&$0$&$0$&$0$&$1$&$0$&$1$&$0$&$0$\\\hline
{\bf E}&$0$&$0$&$-1$&$2$&$-1$&$0$&$0$&$-1$&$2$\\\hline
{\bf F}&$\frac{3 \left(w_m+1\right)}{\alpha }$&$0$&$\frac{9 \left(1-w_m\right) \left(w_m+1\right){}^2}{2 \alpha ^2 w_m}$&$0$& 
 $w_{\text{eff}}$ & 
 $\frac{3 \left(w_m-3\right) \left(w_m+1\right){}^2+2 \alpha ^2 w_m}{2 \alpha ^2 w_m}$ & $\frac{3
   \left(w_m+1\right){}^2}{\alpha ^2}$ &$ -\frac{9 \left(w_m-1\right) \left(w_m+1\right){}^2}{2 \alpha ^2 w_m}$&$0$
\\\hline
{\bf G}&$\frac{3(1+w_m)}{\beta}$&$0$&$0$&$\frac{18(1-w_m)(1+w_m)^2}{\beta^2(1+3w_m)}$& $w_m$ & $\frac{3 \left(3 w_m-7\right) \left(w_m+1\right){}^2+\beta ^2 \left(3 w_m+1\right)}{\beta ^2 \left(3 w_m+1\right)}$ & $\frac{3
   \left(w_m+1\right){}^2}{\beta ^2}$&$0$ &$ -\frac{18 \left(w_m-1\right) \left(w_m+1\right){}^2}{\beta ^2 \left(3 w_m+1\right)}$\\\hline
{\bf H}&$\frac{3(1+w_m)}{\lambda}$&$\frac{9(1-w^2)}{\lambda^2}$&$0$&$0$&$w_m$&$\frac{\lambda ^2-6 \left(w_m+1\right)}{\lambda ^2}$&$\frac{6 \left(w_m+1\right)}{\lambda ^2}$&$0$&$0$\\\hline
{\bf I}&$\frac{\lambda}{2}$&$\frac{12-\lambda^2}{4}$&$0$&$0$&$\frac{\lambda ^2-6}{6}$&$0$&$1$&$0$&$0$\\\hline
{\bf J$_{\alpha}$}&$\frac{2\left(\alpha^2 -a_3^2\right)}{\alpha  a_2}$&$0$&$-\frac{a_4}{3 \alpha ^2 a_2^2}$&$0$&$-$&$\frac{3 \alpha ^2 a_2^2-4 \left(\alpha ^2-a_3^2\right){}^2+a_4}{3 \alpha ^2 a_2^2}$ &$ \frac{4 \left(\alpha
   ^2-a_3^2\right){}^2}{3 \alpha ^2 a_2^2}$ & $-\frac{a_4}{3 \alpha ^2 a_2^2}$ & $0$\\\hline
{\bf J$_{\beta}$}&$\frac{3  \left(\beta^2 -b_3^2\right)}{\beta  b_2}$&$0$&$0$&$\frac{b_4}{3 \beta ^2
   b_2^2}$&$-$& $\frac{3 \beta ^2 b_2^2-9 \left(\beta ^2-b_3^2\right){}^2+b_4}{3 \beta ^2 b_2^2}$&$\frac{3 \left(\beta
   ^2-b_3^2\right){}^2}{\beta ^2 b_2^2}$ & $0$ &$ -\frac{b_4}{3 \beta ^2 b_2^2}$\\\hline\hline
\end{tabular}
\endgroup

}
\label{tab:table01}
\end{table}
\end{widetext}

\begin{table}[ht]
    \centering
    \caption{ Summary of eigenvalues at critical points}
    \resizebox{0.475\textwidth}{!}{ 
   \renewcommand{\arraystretch}{2}
   \begin{tabular}{lcccccc}\hline\hline
 &
$\lambda_1$&
$\lambda_2$&
$\lambda_3$&
$\lambda_4$\\\hline
{\bf A}&$3$&$-3/2$&$3/2$&$-3w_m$\\\hline
{\bf B}&$2$&$2$&$2$&$-3(w_m+1)$\\\hline
{\bf C}&$0$&$0$&$-3$&$-3(1+w_m)$\\\hline
{\bf D$_\pm$}&$3(1-w_m)$&$-6\pm\sqrt{3}\alpha$&$-6\pm\sqrt{3}\beta$&$6\mp\sqrt{3}\lambda$\\\hline
{\bf E}&$-3$&$0$&$0$&$-3(1+w_m)$\\\hline
{\bf F}&$\frac{3 (\beta-\alpha ) \left(w_m+1\right)}{\alpha }$&$\frac{3 }{4}(w_m-1+\sqrt{\gamma_3})$&$\frac{3 }{4}(w_m-1-\sqrt{\gamma_3})$&$\frac{3 (\alpha -\lambda ) \left(w_m+1\right)}{\alpha }$\\\hline
{\bf G}&$-3(1+w_m)$&$\frac{3 }{4}(w_m-1+\sqrt{\gamma_2})$&$\frac{3 }{4}(w_m-1-\sqrt{\gamma_2})$&$\frac{3( \beta-\alpha ) (w_m+1) }{\beta }$\\\hline
{\bf H}&$\frac{3 (\beta -\lambda ) \left(w_m+1\right)}{\lambda }$&$\frac{3 (\alpha -\lambda ) \left(w_m+1\right)}{\lambda }$&$\frac{3 }{4}(w_m-1+\frac{\sqrt{\gamma_1}}{\lambda})$&$\frac{3 }{4}(w_m-1-\frac{\sqrt{\gamma_1}}{\lambda})$\\\hline
{\bf I}&$\frac{ \lambda  (\alpha -\lambda )}{2}$&$\frac{\lambda  (\beta -\lambda
   )}{2} $&$\frac{\lambda ^2}{4}-3$&$\frac{\left(\lambda ^2-6 w_m-6\right)}{2} $\\[5pt]\hline\hline
\end{tabular}

}
    \label{tab:table02}
\end{table}
\section{\label{sec:Discussion} Discussion}
In the present work we studied the autonomous system for the scalar-tensor model with non-minimal kinetic and Gauss-Bonnet couplings. The GB coupling that has been already considered in \cite{Koivisto:2006ai}  contains two scaling solutions: the one given by {\bf GB5} is stable fixed point por $\beta<\lambda$ and $\sqrt{6(1+w_m)}<\lambda\le \sqrt{\frac{48}{7+9w_m}}(1+w_m)$ and is stable spiral for $\beta<\lambda$ and $\lambda\ge \sqrt{\frac{48}{7+9w_m}}(1+w_m)$.  The point {\bf GB6} is an scaling solution associated with the GB coupling, but can not be realized physically for $w_m$ in the region $0\le w_m\le 1$. The analysis shows that is not possible to simultaneously satisfy the condition $0\le\Omega_m\le 1$ and the stability conditions for the fixed point. However, the point can be physical and saddle point for some regions of the parameters as shown for {\bf GB6}. The system also contains the scalar-field dominated fixed point {\bf GB4} that leads to dark energy solution, which is stable for $\beta<\lambda$ and $\lambda<\sqrt{6(1+w_m)}$. An additional point connected with the GB coupling is given by {\bf GB7}. At the limit of large $\beta$ this point turns into the saddle point $(0,0,1)$ ({\bf GB1}), giving $w_{eff}=-1/3$ which is the limit between the decelerated and accelerated expansion.\\
\noindent For the kinetic coupling case we have found two scaling solutions: the one given by the point {\bf K4}, which has the same characteristics as the point {\bf GB5}, and the new scaling solution associated with the kinetic coupling given by the point {\bf K3}. But as in the case of the point {\bf GB6}, in the cosmological scenario described by this point it is not possible to satisfy the conditions of stability for the physical region $0\le\Omega_m\le1$. However, the physical conditions are satisfied and the point becomes saddle as shown in {\bf K3}.  There is also a scalar field dominated critical point {\bf K5} that leads to dark energy solution with the same properties of the point {\bf GB4}. It's worth noting that if $\lambda=\sqrt{6(1+w_m)}$ for this point, the effective EoS becomes $w_{eff}=w_m$, leading in this way to scaling solution. Nevertheless, for $\lambda=\sqrt{6(1+w_m)}$ one of the eigenvalues becomes zero, and the analysis using central manifold techniques (for the case $w_m=0$) shows that the point is unstable (see Appendix B). In any case the point is saddle since two of the eigenvalues become negative (provided that $0<w_m<1$). 

\noindent In the limit $\lambda\rightarrow 0$ the points {\bf GB4} and {\bf K5}  give the critical point  $(0,3,0)$ which leads to a de Sitter solution. The corresponding eigenvalues take the values $[0,-3,-3(1+w_m)]$, indicating that the critical point is at least saddle, and the zero eigenvalue requires additional analysis to see if the point is stable or not. The result of the numerical analysis indicates that the point remains as a saddle point as shown in Fig. 1. Nevertheless, in absence of the interaction terms ( GB coupling for {\bf GB4} and Kinetic coupling for {\bf K5}) the dynamical system becomes two dimensional and the limit $\lambda\rightarrow 0$ leads to a de Sitter attractor with eigenvalues $[-3,-3(1+w_m)]$. 
There is one more critical point connected with the kinetic coupling given by {\bf K6} which has complicated dependence on $\alpha$, but it can be analyzed for specific values. Thus, if we take $\alpha=8/\sqrt{3}$, it gives $w_{eff}=1/3$ and is stable attractor  mimicking radiation dominated universe. In the large $\alpha$ limit, the critical point takes the value $(0,0,1)$ with eigenvalues $[3,3/2,-3w_m]$, coinciding with the saddle point {\bf K1}. The corresponding effective EoS tends to $w_{eff}\rightarrow 0$, indicating that the scalar field at this limit mimics the pressureless matter dominated universe.\\
When we consider the effect of both couplings, we obtain the same critical points discussed for the GB and kinetic couplings with the same physical consequences, and additionally, we obtained the two new critical points {\bf C} and {\bf E} dominated by the non-minimal kinetic and GB couplings. Note that the point {\bf E} coincides with the point {\bf C} for $k=-1$, and both points have the same stability properties and lead to a de Sitter universe. According to the eigenvalues the points are marginally stable, but the central manifold analysis shows that the points remain saddle. To analyze the stability around the point {\bf E}, we can simplify the system by neglecting the potential term. In absence of potential the dimension of the system reduces to three with eigenvalues $[-3,0,-3(1+w_m)]$, and according to the numerical analysis, the point is stable as shown in Fig. 6.  Thus, in absence of the scalar potential, the non-minimal kinetic and GB terms drive the system towards a de Sitter attractor.\\
In the present work we considered the EoS for the matter component only in the range $0\le w_m\le 1$, since we were interested in the role of the kinetic and GB couplings in modeling the dark energy. Nevertheless, the same analysis could be extended to more exotic matter energy densities with negative $w_m$. We characterized the potential, the kinetic and the GB couplings by exponential functions, which are appealing from the point of view of string theory, and show a rich phase structutre. \\
\section{Acknowledments}
This work was supported by Universidad del Valle under project CI 7987.
\appendix
\section{Explicit dynamical system}
In this appendix we will consider some technical details relevant in solving the cosmological equations. This appendix concerns only the dynamics of the model specified by equations  (\ref{eq21}-\ref{eq26}). Employing the dimensionless variables (\ref{eq15}) we can write the system (\ref{eq10},\ref{eq12}, \ref{eq13}) to derive the explicit evolution equation for each individual variable.
\begin{widetext}
\begin{equation}\label{eqA1}
\begin{aligned}
x'=&-A x\Big(9 g^2 \big(3 \omega _m+\beta  x-1\big)+3 g \big(k \big(27 \omega _m-\alpha\, x+6\beta  x-9\big)+x^2 \big(9 \omega _m-17\big)+3 \big((y-3) \omega _m+y-1\big)\\
&+2
   \beta  x^3+3 \lambda  x y\big)+2 \big(3 k^2 \big(9 \omega _m+\alpha  x-3\big)+k
   \big(6 x^2 \big(3 \omega _m-4\big)+9 \big((y-3) \omega _m+y-1\big)+2 \alpha 
   x^3+3 x (\alpha +\lambda  y)\big)\\
&+3 x \big(x^3 \big(\omega _m-1\big)+x\big((y-3) \omega _m+y+3\big)-\lambda  y\big)\big)\Big)\\
y'=&A y
 \Big(-9 g^2 (\lambda  x-2)-6 g \big(k \big(6 \omega _m-\alpha x+2 \beta  x+2 \lambda 
   x-6\big)+x \big(x \big(6 \omega _m-2\big)+ 2 x^2 (\beta -\lambda )+\lambda 
   y\big)\big)\\
&+4 \big(-3 k^2 \big(3 \omega _m+\lambda  x-3\big) +k \big(-12 x^2
   \big(\omega _m-1\big)-3 (y-3) \big(\omega _m+1\big)+x^3 (\lambda -2 \alpha
   )-\lambda  x (2 y+3)\big)\\
&-3 x^2 \big(x^2 \big(\omega _m-1\big)   +(y-3)
   \big(\omega _m+1\big)+\lambda  x\big)\big)\Big)\\
   g'=&-A g \Big(9 g^2+3 \omega _m \big(3 g+2 k-2 x^2\big) \big(3 g+3 k+x^2+y-3\big)+3 g \big(k
   (\alpha  x-2 \beta  x+3)+2 \beta  x^3-13 x^2+\lambda  x y+3 y-3\big)\\
&+2 \big(3 k^2
   (x (\alpha -2 \beta )+3)+k \big(2 x^3 (\beta -\alpha )+3 x (\alpha -2 \beta )+y
   (3-\lambda  x)+9\big)+3 x \big(x \big(x^2-2 \beta  x-y+9\big)-\lambda 
   y\big)\big)\Big)\\
   k'=&A k\Big(9 g^2 x (\alpha -2 \beta )-6 (3 g+4 k) \omega _m \big(3 g+3 k+x^2+y-3\big)+6 g \big(k
   (2 \alpha  x-4 \beta  x+3)-2 \alpha  x^3+15 x^2\\
&-2 \lambda  x y-3 y+3\big)- 4
   \big(\alpha  (k-3) x^3-6 (2 k-3) x^2+(k-3) \lambda  x y+6 k y\big)\Big)
\end{aligned}
\end{equation}
\end{widetext}
where $A$ is given by
\begin{equation}
A=\frac{1}{9 g^2-4 x^2 (3 g+k-3)+12 g k+12 k (k+1)}
\end{equation}
and the slow-roll parameter $\epsilon$
\begin{widetext}
\begin{equation}\label{eqA2}
\begin{aligned}
   \epsilon=&-A \Big(
3 g (k (2 \alpha  x-\beta  x+6 \;\omega_m\;-6)+x (2 x (\alpha  x+3 \;\omega_m\;-1)+\lambda  y))-9 g^2+2(9 k^2 (\;\omega_m\;-1)\\
&+k
  (2 \beta  x^3+12 x^2 (\;\omega_m\;-1)+2 \lambda  x y+3 (y-3) (\;\omega_m\;+1))+3 x^2(x^2 (\;\omega_m\;-1)+(y-3)
   (\;\omega_m\;+1)))
\Big)
\end{aligned}
\end{equation}
\end{widetext}

The tables \ref{tab:table01} and \ref{tab:table02}  summarizes the critical points and eigenvalues for the full model. The fixed points $J_{\alpha}$, $J_{\beta}$ correspond to the points {\bf J} ($J_{\alpha}$, $J_{\beta}$)
 which have complicated dependence on $\alpha$ and $\beta$.
\begin{equation}\label{eq100}
J_{\alpha}= \left(\frac{2\left(\alpha^2 -a_3^2\right)}{\alpha  a_2},0,-\frac{a_4}{3 \alpha ^2 a_2^2},0\right)
\end{equation}
 Where 
 \begin{equation}
 \begin{aligned}
 a_1=&\sqrt{\alpha ^2 \left(8 \alpha ^4-183 \alpha ^2+1152\right)}\\ 
 a_2=&\sqrt[3]{-3 \alpha ^2+a_1+64}\\ a_3=&\sqrt{\frac{a_2^2+4 a_2+16}{2}}\\
 a_4=&4 \alpha ^4-\left(a_2
   \left(7 a_2+19\right)+88\right) \alpha ^2+96 a_3^2+a_1 \left(a_2+8\right)
\end{aligned}
\end{equation}
\begin{equation}\label{eq101}
J_{\beta}=\left(\frac{3  \left(\beta^2 -b_3^2\right)}{\beta  b_2},0,0,\frac{b_4}{3 \beta ^2
   b_2^2}\right)
\end{equation}
Where:
\begin{equation}
\begin{aligned}
b_1=&\sqrt{\beta ^2 \left(9 \beta ^4-264 \beta ^2+2000\right)}\\ b_2=&\sqrt[3]{-54 \beta ^2+9 b_1+1000}\\
b_3=&\sqrt{b_2^2+10 b_2+100}/3\\
b_4=&\beta ^2 \left(-9
   \beta ^2+b_2 \left(5 b_2+26\right)+320\right)-300 b_3^2\\
   &-b_1 \left(b_2+20\right)
\end{aligned}
\end{equation}
\section{Centre manifold method}
In the case of zero eigenvalues the linear theory fails to provide information on the stability of the critical point. If the real part of the eigenvalues are less or equal to cero, it is possible to apply centre manifold method. The main aim of the centre manifold is to reduce the dimensionality of the system near that point so that stability of the reduced system can be investigated. The stability of the reduced system determines the stability of the system at that point.\cite{Boehmer:2011tp,wiggins2003introduction,carr1981applications}.\\

In order to determine the stability of the point {\bf C}:  $(x,y,k,g)=(0,3(1+k),k,-2k)$, which leads to a de Sitter universe, it is necessary to perform a coordinate axes transformation, since the centre manifold method applies only at the origin. The transformation takes the form $(x,u,k,v)=(0,y-3(1+k),k,g+2k,)$. In this reference system the coordiantes of the critical point are $(x,u,v,k)=(0,0,0,0)$. Performing linear analysis, the new eigenvalues take the form $[0,\frac{3 (k+1)}{2 k+1},-3]$. In this case in order to establish the conditions of the method, it is necessary that $\frac{3 (k+1)}{2 k+1}\leq 0$. In the critical case $k=-1$  there are two zero eigenvalues  and one negative  eigenvalue. After straightforward calculations the parametric equation of the centre manifold becomes
\begin{equation}
\begin{aligned}
v=\frac{2 \alpha x^3}{3}+\frac{4 x^2}{3}+O(x^4)
\end{aligned}
\end{equation}
The reduced system on the centre manifold is the following
\begin{equation}
\begin{aligned}
x'&=3 x^3-\frac{u x^3}{2}\\
u'&=u^2+u x^2   
\end{aligned}
\end{equation}
Which results in an unstable critical point, and more precisely, in saddle point.\\ 

Another interesting case appears in the point {\bf I} when $\lambda=\sqrt{6}$, which leads to one zero eigenvalue (assuming $w_m=0$). Performing the coordinate transformation $(u,v,k,g)=(x-\lambda/2,(12-\lambda^2)/4,k,g)$, the parametric equations of the centre manifold become:
\begin{equation}
\begin{aligned}
v&=u^2\\
k&=0\\
g&=0\\
\end{aligned}
\end{equation}
And the reduced system takes the from
\begin{equation}
\begin{aligned}
u'=\sqrt{6} u^2
\end{aligned}
\end{equation}
Signaling that the system is unstable. There are other cases, but have not been considered because they are outside the region of interest or do not meet the conditions of the method.
\bibliographystyle{apsrev4-1}
\bibliography{DarkEnergy}

\end{document}